\documentclass[onefignum,onetabnum]{siamonline190516}
\usepackage{hyperref}
\usepackage{pifont}
\usepackage{amsmath}
\usepackage{graphicx}
\usepackage{tikz}
\usetikzlibrary{chains}
\usepackage{wrapfig}
\usepackage{xfrac}
\usepackage{amsfonts,amssymb,amscd}
\usepackage{mathtools}
\usepackage{url}
\usepackage{bookmark}
\usepackage{stmaryrd}
\usepackage{multirow}
\usepackage{epstopdf}
\usepackage{algorithmic}
\usepackage{verbatim,color,colortbl}
\usepackage{xcolor}
\usepackage[super]{nth}

\newcommand{\etal}{\emph{et al.} }
\newcommand{\eg}{\emph{e.g.}}
\newcommand{\ie}{\emph{i.e.}}
\DeclareMathOperator*{\EE}{\mathbb{E}}
\DeclareMathOperator{\Tr}{Tr}
\DeclareMathOperator{\MSE}{MSE}

\newcommand\abs[1]{\left\lvert#1\right\rvert}

\def\bR{{\mathbb {R}}}

\def\bN{{\mathbb {N}}}
\def\bZ{{\mathbb {Z}}}

\def\bn{{\mathbf{n}}}

\def\bz{{\mathbf{z}}}

\def\0{{\mathbf{0}}}
\def\1{{\mathbf{1}}}

\def\bb{{\mathbf{b}}}

\def\bx{{\mathbf{x}}}
\def\by{{\mathbf{y}}}
\def\BI{{\mathbf{I}}}

\def\BA{{\mathbf{A}}}

\def\BU{{\mathbf{U}}}

\def\BM{{\mathbf{M}}}

\def\BP{{\mathbf{P}}}
\def\BR{{\mathbf{R}}}

\newcommand{\bbra}[1]{\left\llbracket{#1}\right\rrbracket}
\newcommand{\floor}[1]{\left\lfloor{#1}\right\rfloor}
\ifpdf
\DeclareGraphicsExtensions{.eps,.pdf,.png,.jpg}
\else
\DeclareGraphicsExtensions{.eps}
\fi

\usepackage{enumitem}
\setlist[enumerate]{leftmargin=.5in}
\setlist[itemize]{leftmargin=.5in}

\newsiamremark{remark}{Remark}
\newsiamremark{hypothesis}{Hypothesis}
\crefname{hypothesis}{Hypothesis}{Hypotheses}
\newsiamthm{claim}{Claim}
\newsiamthm{example}{Example}
\headers{Patch-Based Holographic Image Sensing}{A.~M.~Bruckstein, M.~F.~Ezerman, A.~A.~Fahreza, and S.~Ling}
\title{Patch-Based Holographic Image Sensing
\thanks{Submitted to the editors DATE. The corresponding author is M.~F.~Ezerman.
\funding{This research is supported by National Research Foundation, Singapore, and Israel Science Foundation under their joint program NRF2015-NRF-ISF001-2597. It is also supported by Nanyang Technological University, Grant Number M4080456.}}}

\author{Alfred Marcel Bruckstein\thanks{Department of Computer Science, Technion, Israel Institute of Technology, Haifa 32000, Israel. (\email{freddy@cs.technion.ac.il}) \and School of Physical and Mathematical Sciences, Nanyang Technological University, 21 Nanyang Link, Singapore 637371. (\email{ambruckstein@ntu.edu.sg}).}
\and Martianus Frederic Ezerman\thanks{School of Physical and Mathematical Sciences, Nanyang Technological University, 21 Nanyang Link, Singapore 637371.
(\email{fredezerman@ntu.edu.sg}, \email{adamas@ntu.edu.sg}, \email{lingsan@ntu.edu.sg}).}
\and \linebreak Adamas Aqsa Fahreza\footnotemark[3] \and San Ling\footnotemark[3]}

\usepackage{amsopn}
\DeclareMathOperator{\diag}{diag}

\ifpdf
\hypersetup{
  pdftitle={Patch-Based Holographic Image Sensing},
  pdfauthor={A.~M.~Bruckstein, M.~F.~Ezerman, A.~A.~Fahreza, and S.~Ling}
}
\fi

\begin{document}

\maketitle

\begin{abstract}
Holographic representations of data enable distributed storage with progressive refinement when the stored packets of data are made available in any arbitrary order. In this paper, we propose and test patch-based transform coding holographic sensing of image data. Our proposal is optimized for progressive recovery under random order of retrieval of the stored data. The coding of the image patches relies on the design of distributed projections ensuring best image recovery, in terms of the $\ell_2$ norm, at each retrieval stage. The performance depends only on the number of data packets that has been retrieved thus far. Several possible options to enhance the quality of the recovery while changing the size and number of data packets are discussed and tested. This leads us to examine several interesting bit-allocation and rate-distortion trade offs, highlighted for a set of natural images with ensemble estimated statistical properties.
\end{abstract}



\section{Introduction}\label{sec:intro}

Holographic sensing and representations aim to capture and describe signals, images, videos and other information sources in a way that enables their recovery at various levels of resolution. The quality of the recovered data is dependent only on the size of the data portion that is available. Often, a mere preview of the visual information at some crude resolution level is sufficient to decide whether one needs to have the full detailed description. In such cases, holographic storage of the data is clearly beneficial. 

Holographic representations of images were first proposed by Bruckstein, Holt, and Netravali in~\cite{Bruckstein1998}. Their ideas exploited the redundancy in images, based on either subsampling strategies or on Fourier transforms using a random phase mask, to ensure diffusion of low frequency information over all portions of the data. These ideas were then tested on individual images, but no theoretical evaluations on the expected achievable quality in the image recovery process were provided. 

Several follow-up works, \eg, \cite{Dovgard2004,Dolev2010,Dolev2012}, came up with improved transforms for more efficient holographic encoding processes. Another approach, given in~\cite{BHN00}, proposed jittered quantization to achieve better precision with multiple instances of \emph{compressed} image data exploiting differently quantized transform coefficients. A subsampling method for generating holographic data streams that ensure a uniform spatial spread of the sampled locations using low discrepancy sampling patterns can be found in~\cite{Bruckstein2005}. 

In information theory, the issue of multiple representations has often been raised. Notable among them are the works of Goyal \etal in~\cite{GKAV98} and of Kutyniok \etal in~\cite{Kutyniok2009}. They dealt with coding into packets of data when communication processes may drop some of the packets in unpredictable ways. An earlier work~\cite{Vaish93} of Vaishampayan discussed multiple representations of a random variable via shifted or staggered quantization tables, a very nice idea resulting in better data fidelity as more and more quantized data are provided. Servetto \etal continued on this trend of ideas in~\cite{Servetto2000} by proposing some high-performance wavelet-based multiple description image coding algorithms.

The insight in multiple description coding (MDC) is to encode an information source into multiple packets which are then sent through parallel channels, each of which is \emph{errorless} and \emph{bandwidth constrained}. There are a number of receivers that, upon receiving a subset of the packets, are tasked with the reconstruction. Most treatments in the literature focus on providing optimized performance in dealing with packet erasures, in various channel models, via the formalism of tight frames. 
	
A general problem in MDC is to characterize the {\it achievable rate region} under specific distortion constraints on chosen reconstruction models. Detailed investigations can be found, \eg, in the works by Goyal \etal in \cite{Goyal2001}, by Venkataramani \etal in \cite{Venkataramani2003}, and in the two part works \cite{Pradhan2004} and \cite{Puri2005} of Pradhan, Puri, and Ramchandran. Reaching their clearly stated performance predictions, however, remains an unattainable practical goal. This is mostly due to two factors. First, designing optimally suited tight frames for arbitrary numbers of erasures is highly nontrivial, although a clearer picture is available for $1$ or $2$ erasures. This fact is also evident in \cite{Kutyniok2009}. Second, attempts to establish multi-erasure information theoretic achievability results have not been successful.

Recently we provided an in-depth analysis of a holographic sensing paradigm, in the classical setting of Wiener filtering, on general random vector data in~\cite{Bruckstein2019}. In the proposal, probings are performed by projection operators that eventually enable graceful successive refinement of the random vectors. The probing projection designs ensure the nice property that only the \emph{number of available probings} determines the quality of the recovery. The \emph{order} of their arrival is irrelevant. 

Another interesting approach to holographic coding of data combines standard data compression procedures with shifts in the image plane to obtain a set of compressed versions of images that form the distributed packets of data~\cite{DB19}. In this case, again, random combinations of the compressed versions will yield improvement in the quality of the decompressed images, depending on the number of versions available.

In this paper, we apply the paradigm in~\cite{Bruckstein2019} on images, seen as ensembles of smaller patches which are interpreted as realizations of a random process, in place of random vectors. In this important application, instead of merely repeating the relevant process in \cite{Bruckstein2019}, we carefully and deeply exploit special statistical properties of image data vectors. Furthermore, we formulate explicit probing policies guided by either estimated or modeled statistical image properties. We solve the problem of optimal probing allocations \emph{when all packets are available}, and then discuss on how to distribute them so as to achieve a graceful and smooth degradation \emph{when the packet numbers are decreased in random orderings}. Our strategies fulfil the goals of holographic representations on images.
	
We design the probing projections, based on the statistics derived from a given data set of images, in Sections~\ref{sec:model} and~\ref{sec:design}. The probings are guided either by the joint statistical properties of all the images in the set or by the properties of the patches of the image that is being considered for distributed holographic storage or communication. We ran our projection design process, along with the associated optimal bit allocation rule, on a set of natural images from various sources, described in Section~\ref{sec:database}, and tested the image retrieval quality for various levels of noise, assumed to affect the holographic data acquisition process. The lowest noise level was actually modeling relatively noise-free projections. The higher levels were considerably deteriorating the projections that provide the distributed packets of images. The implemented system allows for modifications on the size of the image patches, the computation of the statistics for the patch ensemble to be encoded, and the number of holographic projections desired. Then, based on the ensemble's autocorrelation's spectrum of eigenvalues, we derive the projection operators and compute the predicted performance curves for recovery from any number of projections packets selected in either random or incremental orderings. The theoretical predictions are then compared to actual image recovery performance. We describe the implementation routines and report the outcomes in Sections \ref{sec:comp} and \ref{sec:control}.

Our present approach, while suboptimal, has excellent properties. Its reliance on straightforward projection operators makes it simple, easy to understand. It yields clearly stated and comprehensible performance results. It provides a readily implemented scheme for holographic representations of images in the presence of noise. In terms of comparison to prior frame-based proposals, we have been unable to find comparable implementation routines that can be adapted with reasonable efforts to measure our approach against. We share a basic implementation software publicly to pave way for comparative analysis in the future.

The following notational conventions are used throughout. Let $0 < k < \ell $ be integers. Denote by $\bbra{\ell}$ the set $\{1,2,\ldots,\ell\}$ and by $\bbra{k,\ell}$ the set $\{k,k+1,\ldots,\ell\}$. Let $\bZ$, $\bN$ and $\bR$ denote, respectively, the set of integers, the set of positive integers, and the field of real numbers. Vectors are expressed as columns and denoted by bold lowercase letters. Matrices are represented by either bold uppercase letters or upper Greek symbols. An $n \times n$ diagonal matrix with diagonal entries $v_j: j \in \bbra{n}$ is denoted by $\diag(v_1,v_2,\ldots,v_n)$. The identity matrix is $\BI$ or $\BI_n$ if the dimension $n$ is important. Concatenation of vectors or matrices is signified by the symbol $|$ between the components. The transpose of a matrix $\BA$ is $\BA^{\top}$.

\section{Our Distributed Sensing Model and Recovery Procedure}\label{sec:model}

The data of interest in our investigation is a random process, \ie, a set of random column vectors $\{\bx_{\omega} : \omega \in \Omega \}$ indexed by a set $\Omega \subset \bN$. The process, of dimension $M$ over $\bR$, is characterized by its expectation $\overline{\bx}$ and its $M \times M$ autocorrelation matrix
\begin{equation}\label{eq:Rxx}
\BR_{xx}:=\EE (\bx_{\omega}-\overline{\bx}) (\bx_{\omega}-\overline{\bx})^{\top}.
\end{equation}
It is assumed that $\overline{\bx}$ is stored and available, and we deal with the centered case of $\overline{\bx} = \0$ and consider the diagonalization of $\BR_{xx}$, which is a positive definite symmetric matrix, with spectral decomposition given by $\BR_{xx} = \Psi \Lambda \Psi^{\top}$. Here $\Lambda= \diag(\lambda_1,\lambda_2,\ldots,\lambda_M)$, with $\lambda_1 \geq \lambda_2 \geq \ldots \geq \lambda_M > 0$, is a diagonal matrix and 
$\Psi = \left[\Psi_1 | \Psi_2 | \ldots | \Psi_M \right]$ is a unitary matrix whose columns, $\Psi_j$ for $j \in \bbra{M}$, are the eigenvectors corresponding to the respective eigenvalues $\lambda_j$. 

Let $m < M$ be a fixed integer. We denote by $\BP_{k}$ the projection matrix $\BU_k \BU_k^{\top}$, where $\BU_k$ is the $M \times m$ matrix that displays the $k^\text{th}$ element of an orthonormal basis of the $m$-dimensional subspace $V_k$ of $\bR^{M}$ that $\BP_{k}$ projects onto. The assumed model to sense realizations of the process $\{ \bx_{\omega} : \omega \in \Omega\}$ is the set of $N \in \bN$ vectors 
\[
\{\bz_k := \BU_k^{\top} \Psi^{\top} \bx_{\omega} + \bn_k : k \in \bbra{N}\}.
\]
Here $\{\Psi^{\top} \bx_{\omega}\}$ consists of column vectors with $M$ entries with autocorrelation
\[
\EE [\Psi^{\top} \bx_{\omega} \bx_{\omega}^{\top} \Psi] 
= \Psi^{\top} \BR_{xx} \Psi = \Lambda. 
\]
Hence, the realizations of the random process $\{\Psi^{\top} \bx_{\omega}\}$ are vectors with uncorrelated entries having variances $\lambda_1, \lambda_2, \ldots, \lambda_M$. The noise vectors $\bn_k$ affecting the probings are realizations of a random process with zero mean and autocorrelation $\sigma_n^2 \BI_{m}$ and is assumed to be independent of $\bx_{\omega}$. This model of sensing implements, first, a \emph{decorrelating transform coding} on the data vector $\bx_{\omega}$, then projects $\Psi^{\top} \bx_{\omega}$ via $\BU_k^{\top}$ onto an $m$-dimensional subspace of $\bR^M$ for a fixed $m <M$. 

There are many possible ways to define the projections, \eg, $\BU_k$ can be any $M \times m$ matrix whose $m$ column vectors, each having $M$ entries, are orthogonal. We choose $\BU_k$ to have columns selected from the standard basis
\[
\Big\{ \bb_1:= 
\begin{bmatrix}
1 & 0 & 0 &\ldots & 0
\end{bmatrix}^{\top}, 
\bb_2:= 
\begin{bmatrix}
0 & 1 & 0 & \ldots & 0
\end{bmatrix}^{\top},
\ldots,
\bb_M:= 
\begin{bmatrix}
0 & \ldots & 0 & 1
\end{bmatrix}^{\top}
\Big\}
\]
of $\bR^M$. In this case the projection $\BP_{k}$ is a diagonal $M \times M$ matrix with $m$ ones at specific locations on the diagonal as determined by the selection of the standard basis vectors in $\BU_k$. Thus, our sensing model provides $N$ vectors, each of dimension $m$, that convey the sensing information on the vectors $\Psi^{\top} \bx_{\omega}$. 

When all $N$ vectors $\{\bz_k : k \in \bbra{N} \}$ are available, we can recover $\bx_{\omega}$ by computing the best linear estimate under the given probing model at the noise level $\sigma_n^2$ in all of the observations. This recovery process provides the expected least-squared optimal solution for the recovery of the data $\{\bx_{\omega} : \omega \in \Omega\}$ when both the data and the noise, seen as random processes, are Gaussian.

The best estimator for $\bx_{\omega}$, given any set of $\ell$ out of $N$ packets 
\begin{equation}\label{eq:sensed}
\Big\{ \bz_{k_j} = \BU_{k_j} \Psi^{\top} \bx_{\omega} + \bn_{k_j} : j \in \bbra{\ell} \Big\}
\end{equation}
is expressed, in the classical Wiener filter, by 
\begin{equation}\label{eq:recovered}
\widehat{\bx}_{\omega} := \BR_{xz_{\mbox{combi}}} \left(\BR_{z_{\mbox{combi}} z_{\mbox{combi}}}\right)^{-1} \bz_{\mbox{combi}}
\end{equation}
from the observation
\[
\bz_{\mbox{combi}}=
\underbrace{\begin{bmatrix}
	\bz_{k_1}\\
	\bz_{k_2}\\
	\ldots\\
	\bz_{k_{\ell}}
	\end{bmatrix}}_{(\ell \cdot m) \times 1}
=\underbrace{\begin{bmatrix}
	\BU_{k_1}^{\top}\\
	\BU_{k_2}^{\top}\\
	\ldots\\
	\BU_{k_{\ell}}^{\top}\\
	\end{bmatrix}}_{(\ell \cdot m) \times M} \Psi^{\top} \bx_{\omega} + 
\underbrace{\begin{bmatrix}
	\bn_{k_1}\\
	\bn_{k_2}\\
	\ldots\\
	\bn_{k_{\ell}}\\
	\end{bmatrix}}_{(\ell \cdot m) \times 1}.
\]
The matrices $\BR_{xz_{\mbox{combi}}}$ and $\BR_{z_{\mbox{combi}} z_{\mbox{combi}}}$ are readily obtained, respectively, as follows
\begin{align*}
\BR_{xz_{\mbox{combi}}} &= \EE \left[\bx_{\omega} \bz_{\mbox{combi}}^{\top}\right] = 
\BR_{xx} \left[ \Psi \BU_{k_1} | \Psi \BU_{k_2} | \ldots | \Psi \BU_{k_{\ell}} \right]\\
&= \BR_{xx} \Psi \left[\BU_{k_1} | \BU_{k_2} \ldots | \BU_{k_{\ell}} \right], \mbox{ and}
\\
\BR_{z_{\mbox{combi}} z_{\mbox{combi}}} &= \EE \left[\bz_{\mbox{combi}} \bz_{\mbox{combi}}^{\top}\right] = 
\begin{bmatrix}
\BU_{k_1}^{\top}\\
\BU_{k_2}^{\top}\\
\ldots\\
\BU_{k_{\ell}}^{\top}\\
\end{bmatrix}
\Lambda 
\left[\BU_{k_1} | \BU_{k_2} \ldots | \BU_{k_{\ell}} \right] + \sigma_n^2 \BI_{(\ell \cdot m)}.
\end{align*}
Since $\BR_{xx} = \Psi \Lambda \Psi^{\top}$, the best reconstructed $\bx_{\omega}$ from the given data is therefore 
\begin{equation}\label{eq:bestrec}
\widehat{\bx}_{\omega} = 
\Psi \Lambda \left[\BU_{k_1} | \BU_{k_2} \ldots | \BU_{k_{\ell}} \right]
\left(\begin{bmatrix}
\BU_{k_1}^{\top}\\
\BU_{k_2}^{\top}\\
\ldots\\
\BU_{k_{\ell}}^{\top}\\
\end{bmatrix}
\Lambda 
\left[\BU_{k_1} | \BU_{k_2} \ldots | \BU_{k_{\ell}} \right] + \sigma_n^2 \BI_{(\ell \cdot m)}\right)^{-1} ~ \bz_{\mbox{combi}}.
\end{equation}
The expected squared-error covariance in the reconstruction is therefore given by
\begin{align}\label{eq:Ree}
\BR_{ee} &= \EE \left[\left(\bx_{\omega} - \widehat{\bx}_{\omega}\right) 
\left(\bx_{\omega} - \widehat{\bx}_{\omega} \right)^{\top} \right]
=\left(\BR_{xx}^{-1}+\frac{1}{\sigma_n^2} \Psi 
\left[\BU_{k_1} | \BU_{k_2} | \ldots | \BU_{k_{\ell}}\right] 
\begin{bmatrix}
\BU_{k_1}^{\top}\\
\BU_{k_2}^{\top}\\
\ldots\\
\BU_{k_{\ell}}^{\top}\\
\end{bmatrix}
\Psi^{\top} \right)^{-1}
\\ \notag
& = \left(\Psi \Lambda^{-1} \Psi^{\top}+ \frac{1}{\sigma_n^2} 
\Psi \sum_{j =1}^{\ell} \BP_{k_j} \Psi^{\top}\right)^{-1}=
\Psi \left(\Lambda^{-1}+\frac{1}{\sigma_n^2} \sum_{j =1}^{\ell} \BP_{k_j}\right)^{-1} \Psi^{\top},
\end{align}
where $\BP_{k_j} = \BU_{k_j} \BU_{k_j}^{\top}$. Our choice of the basis for $\bR^{M}$ implies  
$\BU_{k_j}:=\left[ \bb_{t_1} | \bb_{t_2} | \ldots | \bb_{t_m} \right]$ is a selection of $m$ out of $M$ standard basis vectors of $\bR^{M}$. 
This structure yields an explicit form for the expected mean squared error ($\MSE$) since we have 
\[
\sum_{j =1}^{\ell} \BP_{k_j} = \diag(s_1,s_2,\ldots,s_M),
\]
where $s_i$ is the total number of times that the basis vector $\bb_i$ appears in the combined matrix $\left[\BU_{k_1} | \BU_{k_2} | \ldots | \BU_{k_{\ell}}\right]$.  
Thus, the error covariance in Equation (\ref{eq:Ree}) becomes
\begin{align*}
\BR_{ee} &= \Psi \left(\Lambda^{-1} + \frac{1}{\sigma_n^2} \diag(s_1,s_2,\ldots,s_M) \right)^{-1} \Psi^{\top}\\
&= \Psi ~\diag\left(\frac{\lambda_1}{1+\frac{\lambda_1}{\sigma_n^2} s_1} , 
\frac{\lambda_2}{1+\frac{\lambda_2}{\sigma_n^2} s_2}, \ldots, \frac{\lambda_M}{1+\frac{\lambda_M}{\sigma_n^2} s_M}\right)~\Psi^{\top}.
\end{align*}
The total $\MSE$ in estimating $\bx_{\omega}$ is the trace of the error covariance matrix, \ie, 
\begin{equation}\label{eq:msetotal}
\MSE_{\mbox{total}} = \Tr(\BR_{ee})=\sum_{i=1}^{M} \frac{\lambda_i}{1+\frac{\lambda_i}{\sigma_n^2} s_i},
\end{equation}
where $s_i$ counts the number of times the $i$\textsuperscript{th} entry of $\Psi^{\top} \bx_{\omega}$ has been probed in the sensing process.

In summary, we have explicit expressions for the $\MSE$ incurred in the recovery procedure from any set of combined sensing results. We can therefore use this result to guide the design of the actual sensing process. The sensing process yields $N$ vectors of length $m$ to be distributed as packets of acquired sensing results. As mentioned earlier, the data consist of random vectors of $M$ entries with zero mean and known autocorrelation of a stochastic process. 

We now turn our focus to the design of the $N$ projections $\BP_{1}, \BP_{2}, \ldots, \BP_N$, \ie, on selecting the matrices $\BU_{1}, \BU_{2}, \ldots, \BU_N$ that satisfy the following holographic criteria. First, the distributed restoration must have the progressive refinement property as more packets are utilized. Second, the process must be smooth in the sense that it achieves maximal uniformity and low $\MSE$ estimation when any $\ell$ out of $N$ packets of representation are made available for every $\ell \in \bbra{N}$.

\section{The Holographic Sensing Design}\label{sec:design}

This section explicitly describes the steps of the projection design optimization in terms of the packet allocation process to ensure that the recovery process from randomly ordered packets meets the holographic requirements. The design is presented here in three subsections to mirror the progression in our optimization strategies. First, we think of the design as a general \emph{resource allocation problem}, where we are given some number $K$ of probings to allocate. Second, we include several practical considerations. Third, we propose a simple way to build the projection matrices for the actual sensing.

\subsection{The Optimal Probing Trade-Offs}\label{subsec:optprobe}

We regard the problem of sensing design as a resource allocation, \ie, rate distortion trade-off process. We have random vectors $\by_{\omega} = \Psi^{\top} \bx_{\omega} \in \bR^M$ with uncorrelated entries of variances $\lambda_1,\ldots, \lambda_M$ and we can probe batches of $m$ entries, with some small additive independent noise of zero mean and variance $\sigma_n^2$. We want to recover $\bx_{\omega}$, from a certain number of $m$-probes, via the optimal Wiener filter estimation process. The combined total expected mean squared error in estimating the entries of $\bx_{\omega}$ depends on the number of times $s_1,s_2,\ldots,s_M$ each entry is probed since
\begin{equation}\label{eq:MSE}
\MSE\left(\Lambda, \sigma_n^2, \{s_1,s_2,\ldots,s_M\}\right) = 
\sum_{i=1}^{M} \frac{\lambda_i}{1+\frac{\lambda_i}{\sigma_n^2} s_i},
\end{equation}
as we have derived in Equation (\ref{eq:msetotal}). A natural question arises. How should we design the probes so as to obtain the smallest $\MSE$, given that we select $m$ locations to probe with each $\BU_k \BU_k^{\top} = \BP_k$ projection matrix? 

It is clear that if we have $N$ projections and each of them selects $m$ locations, then we have $\sum_{i=1}^M s_i = N \cdot m$. For any subset of size $\ell$ out of the $N$ projections, \ie, for $\BU_{k_1}, \BU_{k_2}, \ldots, \BU_{k_{\ell}}$ where $k_j \in \bbra{N}$, we have a total of $\ell \cdot m$ probings. This leads us to consider the following questions. 
\begin{enumerate}[wide, itemsep=0pt, leftmargin =0pt,widest={{\bf Question 1}}]
\item[{\bf Question $1$}:] What is the lowest $\MSE$ attainable with a given total number of $K$ probings?
\item[{\bf Question $2$}:] What is the optimal distribution of probings, \ie, the allocation of the $\ell \cdot m$ probings in total to the set $\{s_i\}$ of nonnegative integers that yields the lowest $\MSE$?
\end{enumerate}
We are thus led to the following optimization. Let $K$ denote the total number of probings to be suitably distributed into $\ell$ probing projections. Given
$\lambda_1 \geq \lambda_2 \geq \ldots \geq \lambda_M > 0$ and $\sigma_n^2$, minimize, over all suitable $\{s_j : j \in \bbra{M}\}$ allocations, the $\MSE$ in Equation (\ref{eq:MSE}), subject to the requirements that $0 \leq s_j \leq N$ must be an integer for all $j \in \bbra{M}$ and $\sum_{j=1}^M s_j = K$.

To address this resource allocation problem, we begin with the last condition on $s_j$, \ie, we relax the optimization over $\bR$, instead of $\bZ$. To make this relaxation clear, we replace $s_j \in \bZ$ by $\zeta_j \in \bR$. This reduces the problem into a straightforward Lagrangian optimization problem, requiring us to solve
\begin{equation}\label{eq:Lagrange}
\frac{\partial}{\partial \zeta_j} 
\left(
\sum_{j=1}^{M} \frac{\lambda_j}{1+\frac{\lambda_j}{\sigma_n^2} \zeta_j} + \gamma 
\left(K - \sum_{j =1}^M \zeta_j \right)\right)=0.
\end{equation}
Solving it yields, for each $j \in \bbra{M}$,
\[
\frac{\lambda_j^2 \cdot  \sigma_n^2} {\left(\sigma_n^2 + \lambda_j \zeta_j \right)^2} = \gamma 
\iff 
\left(\sigma_n^2 + \lambda_j \zeta_j\right)^2 = \frac{1}{\gamma} \lambda_j^2 \cdot \sigma_n^2,
\]
which implies 
\[
\sigma_n^2 + \lambda_j \zeta_j = \frac{1}{\sqrt{\gamma}} \lambda_j \cdot \sigma_n.
\]
Thus, the solution to the optimization problem is the set 
\begin{equation}\label{eq:SSet}
\bigg\{\zeta_j = \frac{\sigma_n}{\sqrt{\gamma}} - \frac{\sigma_n^2}{\lambda_j} : j \in \bbra{M}\bigg\}. 
\end{equation}
Now, adding the constraint $\sum_{k=1}^M \zeta_k =K$ means
\begin{align*}
\frac{\sigma_n}{\sqrt{\gamma}} \cdot M - \sigma_n^2 \sum_{k=1}^M \frac{1}{\lambda_k} = K 
& \iff  
\frac{1}{\sqrt{\gamma}} = \frac{1}{M \cdot \sigma_n} \left(K + \sigma_n^2 \sum_{k=1}^M \frac{1}{\lambda_k}\right) \\
& \iff 
\frac{1}{\sqrt{\gamma}} = \frac{K}{M \cdot \sigma_n} + \frac{\sigma_n}{M} \sum_{k=1}^M \frac{1}{\lambda_k}.
\end{align*}
Thus, we obtain
\begin{equation}
\label{eq:KSset}
\zeta_j = \frac{K}{M} + \sigma_n^2 \left( \frac{1}{M} \sum_{k =1}^M \frac{1}{\lambda_k} - \frac{1}{\lambda_j}\right),
\end{equation}
which is very similar to bit allocation in transform coding.

The result we have just obtained is quite nice. The $\zeta_j$s have to be distributed about the even distribution of the $K$ probings to the $M$ entities in the probed vector, \ie, in the ratio $\frac{K}{M}$ according to how far their inverse variance is from the average inverse variances of all entries. The distribution, in other words, depends on the {\bf harmonic mean of the variances}. Clearly, this result has not yet ensured that the other conditions imposed on the $s_j$s, namely $s_j$ must be a nonnegative integer, upper bounded by $N$, for each $j$, are satisfied. Looking at the solutions for $\zeta_j$s, however, we realize that the function 
\begin{equation}
\label{eq:rho}
\rho(j) = \frac{1}{M} \sum_{k=1}^M \frac{1}{\lambda_k} - \frac{1}{\lambda_j}
\end{equation}
is a decreasing function of $j$. We therefore need to have 
\begin{align}\label{eq:K1}
\frac{K}{M} + \sigma_n^2 \left( \frac{1}{M} \sum_{k=1}^M \frac{1}{\lambda_k} - \frac{1}{\lambda_M} \right) \geq 0 &\iff 
\frac{K}{M} \geq \sigma_n^2 \left( \frac{1}{\lambda_M} - \frac{1}{M} \sum_{k=1}^M \frac{1}{\lambda_k} \right) \notag\\
& \iff K \geq M \cdot \sigma_n^2 \left( \frac{1}{\lambda_M} - \frac{1}{M} \sum_{k=1}^M \frac{1}{\lambda_k} \right)
\end{align}
to guarantee that $\zeta_M$ and, hence, $\zeta_1, \ldots, \zeta_{M-1}$ will all be nonnegative.
If the above conditions are satisfied, then we will also have 
\begin{equation}\label{eq:K2}
\zeta_1 = \frac{K}{M} + \sigma_n^2 \left( \frac{1}{M} \sum_{k =1}^M \frac{1}{\lambda_k} - \frac{1}{\lambda_1}\right) \leq K. 
\end{equation}
To see this, we notice that  
\[
\frac{1}{\lambda_1} + (M-1) \frac{1}{\lambda_M} \geq \sum_{k =1}^M \frac{1}{\lambda_k} 
= \frac{M-1}{M} \sum_{k =1}^M \frac{1}{\lambda_k} + 
\frac{1}{M}
\sum_{k =1}^M \frac{1}{\lambda_k}
\]
implies  
\[
(M-1) \left(\frac{1}{\lambda_M} - \frac{1}{M} \sum_{k =1}^M \frac{1}{\lambda_k}\right) \geq 
\frac{1}{M} \sum_{k =1}^M \frac{1}{\lambda_k} - \frac{1}{\lambda_1}.
\]
Hence, 
\[ 
\frac{1}{\lambda_M} - \frac{1}{M} \sum_{k=1}^M \frac{1}{\lambda_k} \geq \frac{1}{M-1} 
\left(\frac{1}{M} \sum_{k =1}^M \frac{1}{\lambda_k} - \frac{1}{\lambda_1} \right).
\]
Substituting this last expression into the inequality in Equation (\ref{eq:K1}) gives us 
\[
K \geq \frac{M}{M-1} \cdot \sigma_n^2 \left(\frac{1}{M} \sum_{k =1}^M \frac{1}{\lambda_k} - \frac{1}{\lambda_1} \right),
\]
which becomes the inequality in Equation (\ref{eq:K2}) after a simple rearrangement. Thus, we have determined a condition on $K$ to ensure the nonnegativity of all the optimal values assigned to the nonincreasing sequence of real values $\zeta_1, \zeta_2,\ldots, \zeta_m$ whose sum is $K$.

Reflecting on our journey so far, minimizing the total $\MSE$ in estimating $\by_{\omega}$, hence $\bx_{\omega}$, from $K$ probings of its entries is achieved by distributing the probing projections according to the set of integers $s_1,s_2,\ldots,s_M$ determined via 
$\displaystyle{
	\min_{\{s_j\}} \sum_{j =1}^M \frac{\lambda_j}{1+ \frac{\lambda_j}{\sigma_n^2} s_j}}$ such that $\displaystyle{\sum_{j =1}^M s_j = K}$. The minimal value is attained when the integer $s_j$, for each $j \in \bbra{M}$, is ``near'' the real positive value
\begin{equation}\label{eq:zeta}
\zeta_j = \frac{K}{M} + \sigma_n^2 \left(\frac{1}{M} \sum_{k=1}^M \frac{1}{\lambda_k} - \frac{1}{\lambda_j}\right) \in \bR.
\end{equation}
Note that by using the real-valued $\zeta_j$ to compute for $\MSE$ in Equation (\ref{eq:MSE}) we obtain
\begin{align}\label{eq:bestMSE}
\MSE_{\mbox{best}}&=
\sum_{j=1}^M \frac{\lambda_j}{1 + \frac{\lambda_j}{\sigma_n^2} 
	\left(\frac{K}{M} + \sigma_n^2 \left(\frac{1}{M} \sum_{k =1}^M \frac{1}{\lambda_k} - \frac{1}{\lambda_j}\right)\right)}
= \sum_{j=1}^M \frac{\lambda_j}{\frac{\lambda_j}{\sigma_n^2} \frac{K}{M} + 
	\lambda_j \left(\frac{1}{M} \sum_{k=1}^M \frac{1}{\lambda_k}\right)} \notag \\
&= \sum_{j=1}^M \frac{1}{\frac{K}{M \cdot \sigma_n^2} + \frac{1}{M} \sum_{k=1}^M \frac{1}{\lambda_k}} = M \cdot \frac{\sigma_n^2}{\frac{K}{M} + \sigma_n^2 \left(\frac{1}{M} \sum_{k=1}^M \frac{1}{\lambda_k}\right)}.
\end{align} 
Thus, the optimal assignment is achieved when all of the errors in estimating the components of the random vector $\by_{\omega}$ are equalized. This is a very natural and oft-encountered condition in distributed estimation. Keep in mind that this is the case when $K$ is large enough to make all of the $s_j$s nonnegative, \ie, 
$\displaystyle{K \geq M \cdot \sigma_n^2 \left(\frac{1}{\lambda_M} - \frac{1}{M} \sum_{k=1}^M \frac{1}{\lambda_k}\right)}$ from Equation (\ref{eq:K1}). Such a $K$ ensures that the best $\MSE$ in estimating {\it each component} of $\by_{\omega}$ is bounded above by $\lambda_M$, \ie, it stays no more than the smallest variance $\lambda_M$ in the uncorrelated random vector $\by_{\omega}$.

The {\it coding gain}, defined to be the \emph{reduction} in the $\MSE$ value as a consequence of using the optimal $\{\zeta_j\}$ in Equation (\ref{eq:zeta}) instead of $\zeta_j := \frac{K}{M}$ for all $j \in \bbra{M}$, is 
\[
\sum_{j =1}^M \frac{\lambda_j}{1 + \frac{\lambda_j}{\sigma_n^2} \frac{K}{M}} - 
\sum_{j=1}^M \frac{\lambda_j}{1 + \frac{\lambda_j}{\sigma_n^2} 
\left(\frac{K}{M} + \sigma_n^2 \left(\frac{1}{M} \sum_{k =1}^M \frac{1}{\lambda_k} - \frac{1}{\lambda_j}\right)\right)} \geq 0.
\]
The gain is $0$ if and only if $\lambda_1 = \lambda_2 = \ldots = \lambda_M$.

One can plot the function $\displaystyle{\rho(j)=\frac{1}{M} \sum_{k=1}^M \frac{1}{\lambda_k} - \frac{1}{\lambda_j}}$ (see Equation (\ref{eq:rho})), given the $\lambda_j$ values. The curve displaying the allocation of $\zeta_j$ is then influenced by $\frac{K}{M}$, which serves as the constant level around which each $\zeta_j$ lives, starting above it before decreasing as $j$ goes to $M$. The shape is determined by $\sigma_n^2 \cdot \rho(j)$ where $\rho(j)$ depends only on the diagonal entries $\lambda_1,\lambda_2, \ldots, \lambda_M$ of $\Lambda$. A plot of $\rho(j)$ based on actual $\lambda_j$ values calculated from a set of images, alongside the plots of $\zeta_j$ for several values of $\sigma_n^2$, can be found later in Figure~\ref{fig:rho_zeta}.

\subsection{Practical Calibrations}
Given the nonincreasing sequence of variances 
\[
S:=\lambda_1, \lambda_2, \ldots, \lambda_M,
\]
the probing distribution is governed by the function $\rho(j)$ in Equation (\ref{eq:rho}). Letting 
\[
\Delta := \frac{1}{M} \sum_{j =1}^M \frac{1}{\lambda_j},
\]
\ie, $\Delta$ is the {\it harmonic mean} of the elements in $S$, we write $\rho(j) = \Delta - \frac{1}{\lambda_j}$. Together with the assumed noise level and the total number $K$ of probings, $\rho(j)$ provides us with the resource allocation strategy that achieves the best $\MSE$ performance.

In practical situations the sequence $S$ of $\lambda_j$s, the noise's standard deviation $\sigma_n^2$, and the total number of probings $K$ are given to us as system parameters. It may happen, therefore, that $K$ is not large enough or $\sigma_n^2$ not small enough to ensure that all $\zeta_j$s are nonnegative. When this is the case, we shall allocate no probings for a subsequence of small $\lambda_j$s at the tail end of $S$. We accomplish this by repeatedly solving the probing allocation problem for leading subsequences of $S$, say for $\lambda_1, \lambda_2, \ldots \lambda_L$ for an $L < M$. We recalculate $\zeta_1,\zeta_2,\ldots,\zeta_L$ for various $L$ until we arrive at the {\it largest} $L$ that makes
\begin{equation}\label{eq:indexL}
\zeta_{i} := \frac{K}{L} + \sigma_n^2 \left(\frac{1}{L} \sum_{j=1}^L \frac{1}{\lambda_j} - \frac{1}{\lambda_i}\right) > 0 \mbox{ for all } i \in \bbra{L},
\end{equation}
implying $s_{i} \geq 0$, and fix $\zeta_{L+1} = \ldots = \zeta_M=0$. The formula in Equation (\ref{eq:indexL}) is derived by solving the suitably modified optimization problem (see Equation (\ref{eq:Lagrange}))
\begin{align}\label{eq:MSE_L}
\theta\left(\zeta_1,\ldots,\zeta_L,0,\ldots,0\right) &= 
\sum_{j=1}^{M} \frac{\lambda_j}{1+\frac{\lambda_j}{\sigma_n^2} \zeta_j} + \gamma 
\left(\sum_{j =1}^M \zeta_j - K\right) \notag \\
& =
\sum_{j=1}^{L} \frac{\lambda_j}{1+\frac{\lambda_j}{\sigma_n^2} \zeta_j} + 
\sum_{j=L+1}^M \lambda_j + 
\gamma \left(\sum_{j =1}^L \zeta_j - K\right).
\end{align}

The allocation of the $K$ probings is then done to the leading $L$ entries of the vector $\by_{\omega}$ to achieve the goal of having holographic reconstructions for $K = N \cdot m$. In our setup, $N$ is the number of packets created by the projections for each $\by_{\omega}$ and $m$ is the fixed size of each probing packet, \ie, the number of entries in each packet, which is the dimension of the image vector after a projection. These $N$ packets are to be stored or distributed in the environment or sent via some lossy communication channel that drops packets or randomizes their delivery time. 

For practical examples that we encounter in the statistics of natural image patches, the elements in $S$ are highly skewed towards the leading part, \ie, the first variance $\lambda_1$ is, relative to the rest of the entries, much larger. The next few variances tend to be high, then all the remaining variances decrease rapidly as the index goes to $M$. We use the following three modes of determining $s_j$ from $\zeta_j$ for $j \in \bbra{L}$. Their respective advantages become apparent in their actual deployment. 

\begin{enumerate}[wide, itemsep=0pt, leftmargin =0pt,widest={{\bf Mode 3}}]
\item[{\bf Mode $1$}:] As much as possible, this mode assigns $s_j$ to be the closest integer to $\zeta_j$ for $j \in \bbra{M}$. It starts by listing the absolute distances from $\zeta_j$ to its nearest integer $\floor{\zeta_j + \frac{1}{2}}$ and sorts these distances from smallest to largest, keeping tab of the respective indices. Guided by this sorted list, the mode assigns $s_j$ to be the nearest integer to $\zeta_j$, choosing between $\floor{\zeta_j}$ and $\lceil \zeta_j \rceil$, as $i$ goes through the indices in the list. The enforcement decreases in priority, \ie, it may reverse the assignment to keep the condition $\sum_{j =1}^M s_j = K$ satisfied.    
\item[{\bf Mode $2$}:] The idea is to initially assign as $s_j$ the integer part $\floor{\zeta_j}$ of $\zeta_j$, for all $j \in \bbra{M}$, and define an integer 
$\delta_1 : = \sum_{j=1}^M (\zeta_j - \floor{\zeta_j})$. Since $\lambda_1$ has the highest variance coefficient, we assign a new value $s_1 \gets \min\left(N, s_1 + \delta_1\right)$. With the updated $s_1$, let $\delta_2 := K - \sum_{j =1}^M s_j$. If $\delta_2 > 0$, then assign $s_2 \gets \min\left(N, s_2 + \delta_2\right)$. One repeats this process until there is no more left-over quantity to assign.
\item[{\bf Mode $3$}:] This mode simply assigns $s_1=\ldots=s_m=N$ and $s_k=0$ for $k > m$. 
\end{enumerate}

\subsection{Designing the Packet Projections}

After obtaining the optimal allocation that yields the best distribution of the given $K$ probings in terms of minimizing the $\MSE$, \ie, having
\[
s_1 \geq  s_2 \geq \ldots \geq s_L \geq 0 \mbox{ and } s_{L+1}= \ldots = s_{M}=0
\]
for some $1 < L < M$, we need to probe only $L$ first entries of $\by_{\omega}$. Recall that the vector has a given autocorrelation $\BR_{yy} = \Lambda$. We now proceed to designing the $N$ packets, each of size $m$, that will optimally distribute the $K=N \cdot m$  probings with the goal of reaching the best-possible recovery with the desired progressive refinement property. 

For $\ell$ out of the total $N$ packets, the number of available probings is $K_{\ell} := \ell \cdot m$. We combine Equation (\ref{eq:MSE}) and the optimal $\zeta_j$s to infer that the best theoretically possible expected error for $\ell$ packets is given by
\begin{equation}\label{eq:best_ell}
\MSE_{\mbox{best}, \ell} = \sum_{j=1}^L \frac{\lambda_j}{1 + \frac{\lambda_j}{\sigma_n^2} \zeta_j} + \sum_{j=L+1}^M \lambda_j.
\end{equation} 
Given specific $\Lambda$, $m$, and $\sigma_n^2$, we can explicitly compute 
$\MSE_{\mbox{best}, \ell}$ for each $\ell \in \bbra{N}$. Plots for actual images will be given in Section~\ref{sec:comp}.

We shall never attain this optimal performance for every $\ell$-set of projections, since we optimize the allocation of $s_j$s for best recovery when {\it all} packets are available. We should, however, strive to design the $N$ packets of projections in such a way that a random selection of $\ell$ out of $N$ packets, \ie, selecting any one of the $\binom{N}{\ell}$ possible combinations of $\ell$ distinct packets, exhibits \emph{smoothness}. In our context, it means that we recover the vector $\by_{\omega}$ as uniformly well as possible and as close as it can be to the $\MSE_{\mbox{best}, \ell}$ value.

Given the sequence $S$ whose $M$ elements are the diagonal entries of $\Lambda$, $\sigma_n^2$, and $K = N \cdot m$, we want to find the best way in assigning the projections into $N$ boxes satisfying the following conditions.
\begin{enumerate}[wide, itemsep=0pt, leftmargin =0pt,widest={{\bf Condition 1}}]
\item[{\bf Condition $1$}:] The projections sample the $\nth{1}$, the $\nth{2}$, and so on, up to the $L \textsuperscript{th}$ entry of $\by_{\omega}$.
\item[{\bf Condition $2$}:] The $s_j : j \in \bbra{L}$ add up, by design, to a total of $N \cdot m$ projections, \ie, 
\[
s_1 + s_2 + \ldots + s_L = N \cdot m.
\]
\item[{\bf Condition $3$}:] Choices of any $\ell$ out of $N$ boxes, for each $\ell \in \bbra{N}$, yield nearly as good a recovery of $\by_{\omega}$ as possible, without choices that yield significantly bad mean squared errors.
\end{enumerate}

We address the problem as follows. Every projection $\BP_{k} = \BU_k \BU_k^{\top}$ is determined by a set of $m$ basis vectors selected from the standard basis 
$\{\bb_1, \bb_2, \ldots,\bb_M\}$ of $\bR^M$. The numbers $s_1,s_2,\ldots,s_L$ tell us how many times each vector $\bb_i$ will appear in the combined $N$ projections, written as $\sum_{k=1}^N \BP_k$. The key to a good projection is to have as many low-indexed probings as possible since the various projections have different influence on the expected mean squared error of recovery. However, we want every pair of projections to provide similarly good results. The same should hold for any triplet, any quadruplet, and so on, of projections. 

There are several options to distribute the $N \cdot m$ balls of $L$ different colors into $N$ boxes, each with capacity $m$. There are $s_j$ balls of label $j$ for each $j \in \bbra{L}$
\[
\underbrace{1 \qquad 1 \qquad \ldots \qquad 1}_{s_1 \mbox{ balls labelled } 1} \qquad 
\underbrace{2 \qquad 2 \qquad \ldots \qquad 2}_{s_2 \mbox{ balls labelled } 2} \qquad
\ldots \ldots \qquad 
\underbrace{L \qquad L \qquad \ldots \qquad L}_{s_L \mbox{ balls labelled } L}.
\]
Here is a simple yet effective distribution process. What we want is to have, in each of the $N$ boxes, balls with low labels. Such balls contribute the most to the {\bf reduction} of the expected $\MSE$ since they correspond to probing the low-indexed components of $\by_{\omega}$, which have the highest variances. 

We mark the $N$ boxes $B_1$ to $B_N$ and stack them vertically with $B_1$ on top and $B_N$ at the bottom. We then distribute the $s_1$ balls, labelled $1$, one by one to boxes $B_1$ to $B_{s_1}$. As each ball is distributed, move the just-filled box from the very top to the bottom. Once all of these $s_1$ balls are taken care of, the (still empty) top box is now $B_{s_1+1}$ and the bottom box is $B_{s_1}$. Repeat the distribution process on the remaining balls, with the boxes moved from top to bottom as before. Notice that at every instance after the top box is moved to the bottom position, the boxes are ordered in decreasing $\MSE$, with the box on top corresponding to the highest current $\MSE$, if the recovery process is to commence immediately. This way of allocating the probings to the projection boxes will eventually result in a complete allocation of $K = N \cdot m$ probings to the $N$ boxes while making all of these $N$ boxes as similar as possible in terms of their respective $\MSE$ measure of recovery quality. 

\section{The Image Data Set Used for Testing}\label{sec:database}

\begin{figure}[h!]
	\centering
	\begin{tabular}{cc}
		\includegraphics[width=0.4\linewidth]{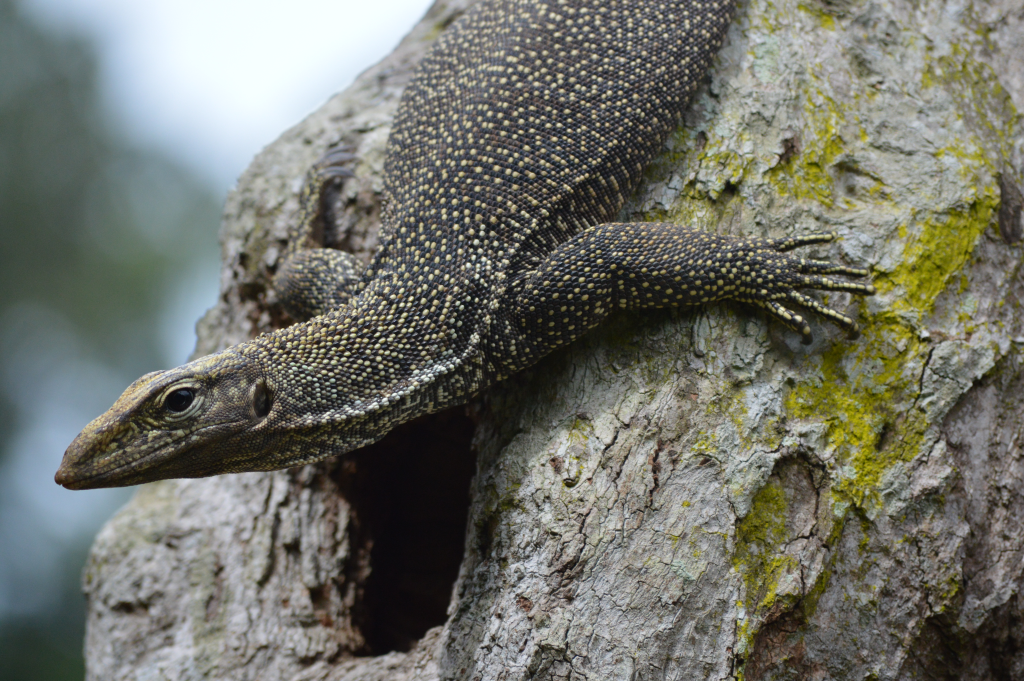} & 
		\includegraphics[width=0.4\linewidth]{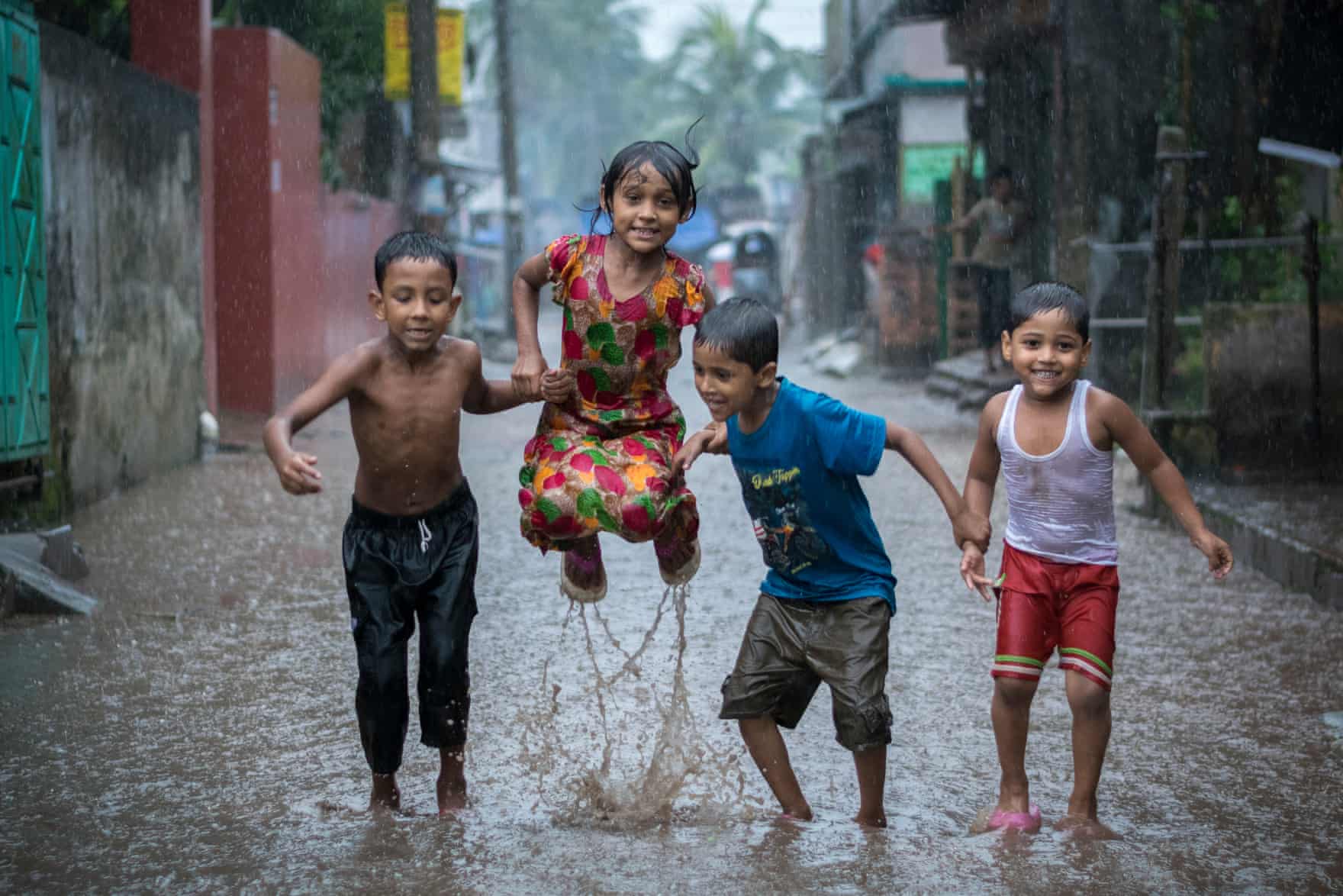} \\
		(a) {\tt dragon} & (b) {\tt flood}\\
		\includegraphics[width=0.4\linewidth]{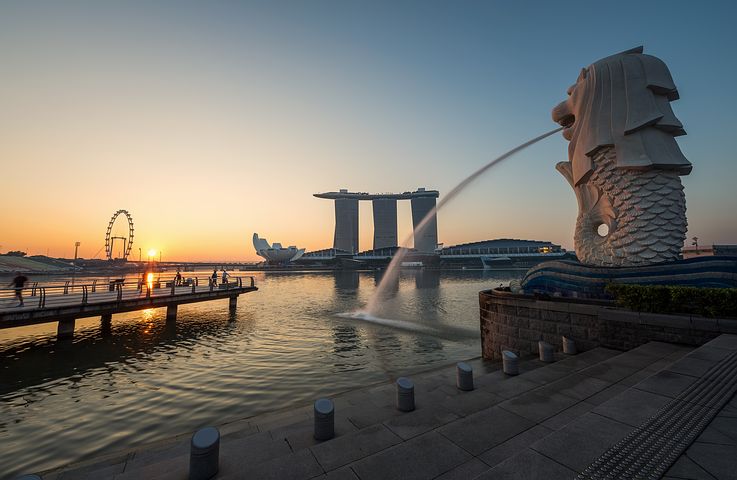} &
		\includegraphics[width=0.4\linewidth]{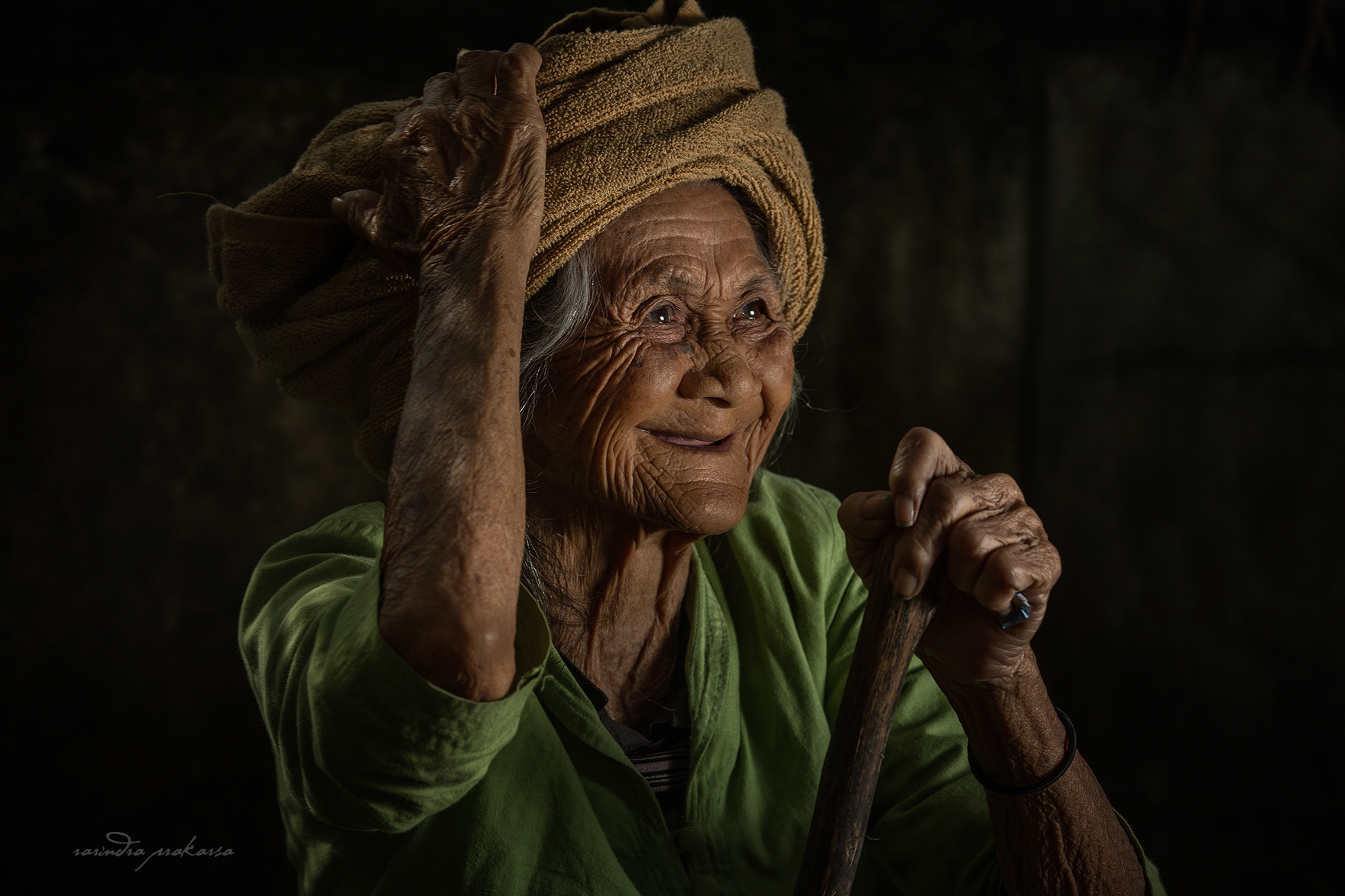} \\
		(c) {\tt merlion} & (d) {\tt oldlady} \\
		\includegraphics[width=0.35\linewidth]{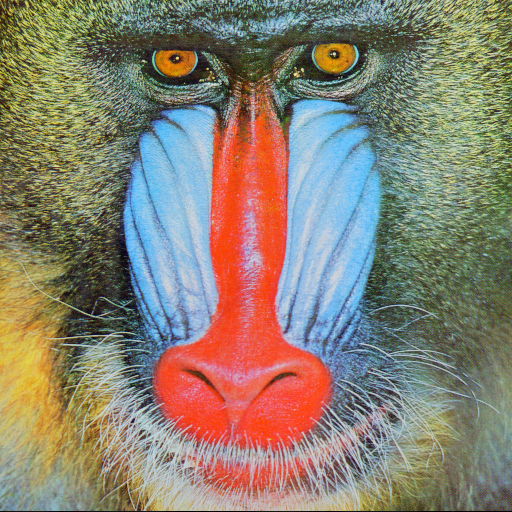} & 
		\includegraphics[width=0.35\linewidth]{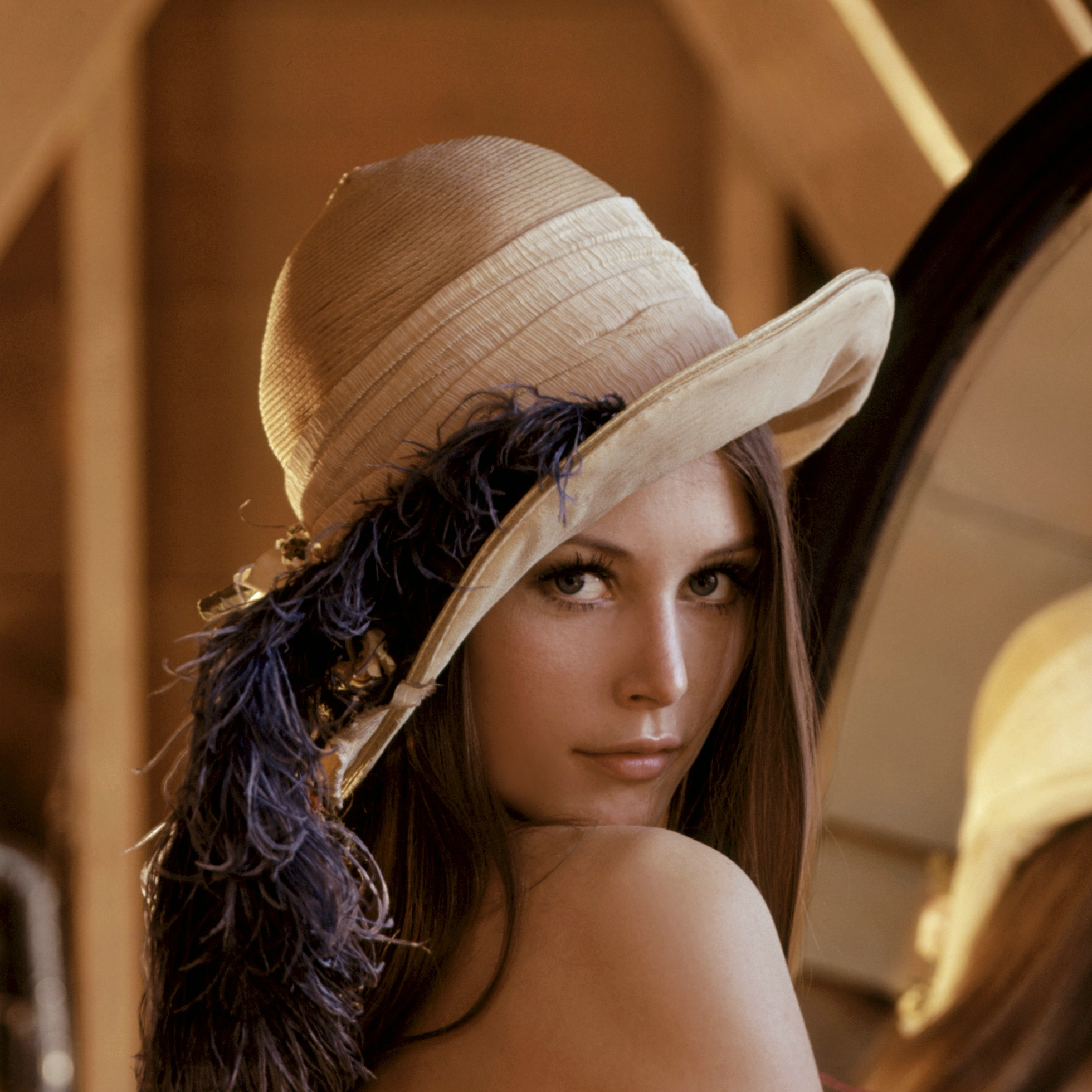} \\
		(e) {\tt mandrill} & (f) {\tt lena}
	\end{tabular}
	\caption{Six of the images from the image data set, to be used as illustrations.}
	\label{fig:images}
\end{figure} 

For a software implementation in Section~\ref{sec:comp}, we use a data set of $49$ images available at~\url{https://github.com/adamasstokhorst/holographic/tree/master/img}. Most of them are either part of the standard images used in image processing or taken from free-to-use online repositories~\url{https://www.pexels.com/} and~\url{https://pixabay.com/}, with no attributions required. Figure \ref{fig:images} presents $6$ of the $49$. 

The image {\tt flood} in Figure \ref{fig:images} (b), was taken by Fardin Oyan. Originally titled {\it Happiness on a Rainy Day}, it won him the Young Environmental Photographer of the Year 2018 award. Two images, namely {\tt oldlady} in Figure \ref{fig:images} (d) and {\tt buffalo} were included in the data set with the permission of their photographer Rarindra Prakarsa. The data set also contains two iconic images, used under the ``fair use principle'' for academic purposes. The award-winning photographer Steve McCurry shot {\it Procession of Nuns} in Rangoon, Burma (now Yangon, Myanmar), in 1994. We call the image {\tt monks}, for brevity. The image {\tt refugee} of a Syrian refugee carying his daughter while crossing the border of Macedonia and Greece in 2015 was taken by the late, decorated photographer, Yannis Behrakis. The images {\tt dragon} in Figure \ref{fig:images} (a) and {\tt baby} belong to the corresponding author.

\section{A Software Implementation}\label{sec:comp}

To demonstrate the efficacy and versatility of holographic sensing on images, we design and thoroughly test an implementation of a distributed image sensing system. The images are then shown to be progressively recovered with the recovery being insensitive to and independent of the order in which the packets became available. Our software is written in {\tt pyton 2.7} with {\tt numpy}~\cite{numpy,Oli06}, Python Imaging Library {\tt PIL}~\cite{PIL}, and {\tt matplotlib}~\cite{matplotlib,Hunter07}, as the minimum required libraries. 

The software implementation routines fall into three types, namely preparatory routines, recovery routines, and performance analysis routines. We welcome readers who are interested to implement the tools for themselves to access the source files at~\url{https://github.com/adamasstokhorst/holographic}. Users can tweak the input parameters to better suit their favourite implementation scenarios.

\subsection{Preparatory Steps}\label{subsec:prep}

In the preprocessing stage, the system analyzes a database of images to obtain the necessary statistical data to use in the sensing design. Dealing with typically large size images, as is customary, we break them up into small patches of $r \times r$ pixels, where $r$ can be $4$, $8$,  or $16$. Each two-dimensional image is row-stacked into vectors of $M=r^2$ entries. Taken sequentially, from all images in the database, these patches form our ensemble of $\Omega$ vectors. These vectors are then analyzed to get their average and, subsequently, the centered autocorrelation. One can easily handle both black-and-white $8$-bit per pixel images and color images, which are simply triplets of RGB $8$-bit per pixel color planes of red, blue, and green added together. The resulting centered ensemble $\{ \bx_{\omega} : \omega \in \Omega \}$ consists of vectors with $M$ entries and zero mean. 

The original value range for the centered images, for $8$-bit quantization, is between $0$ to $255$, with quantization step $1$. The range is then converted in our software to the interval $-1$ to $1$, with the quantization step adjusted accordingly. We set the noise level at $\sigma_n^2 \in \{0.01, 0.25, 0.64, 1.00\}$ for implementation on actual images. Recall that in most of the process, we are looking at $\by_{\omega}$, instead of $\bx_{\omega}$, and, thus, using the diagonalized $\BR_{xx} = \Psi \Lambda \Psi^{\top}$ via $\BR_{yy} = \Lambda$. Once we have the average patch vector and the autocorrelation $\BR_{xx}$ of the ensemble, we perform, for this ensemble, the holographic sensing and recovery processes as described in Sections \ref{sec:model} and \ref{sec:design}. 

\smallskip
\noindent {\it The $\Lambda$ and $\Psi$ Matrices}

\begin{wrapfigure}{r}{0.4\textwidth}
\vspace{-12pt}
\begin{center}
	\includegraphics[width=0.38\textwidth]{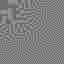}
\end{center}
\caption{A visual presentation of the orthonormal eigenvectors as $8 \times 8$ patches}\label{fig:visual}
\vspace{-5pt}
\end{wrapfigure}

The autocorrelation matrices are of size $M \times M$, for a given $M$. We typically use $M =64$. The standard Singular Value Decomposition function in {\tt numpy} allows for a fast computation of $\Lambda$ and $\Psi$. Figure \ref{fig:visual} displays the orthonormal set of eigenvectors as $8 \times 8$ patches in a combined image as a typical example that can be replicated for other values of $M$. As expected, these eigenpatches look similar to the patches defined by the classical orthonormal basis used in the Discrete Cosine Transform (DCT)~\cite{Ahmed1974}. These are the patches to invoke in reconstructing the images from the data packets that represent the sensed images. 

\begin{figure}[h!]
\centering
\includegraphics[width=0.9\linewidth]{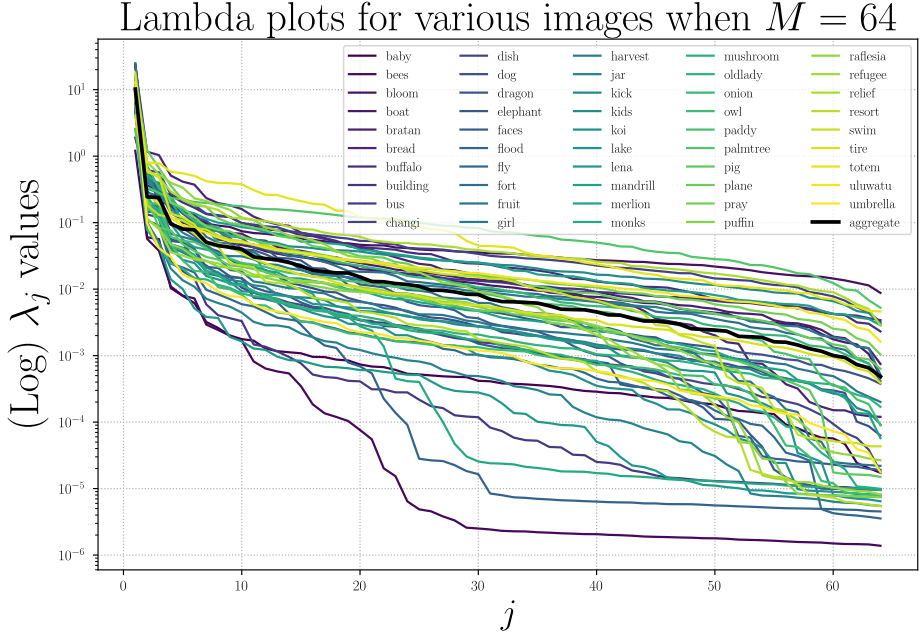} \\
\caption{The $\Lambda$ profiles when $M=64$. The vertical axis is labeled in logarithmic scale.}
\label{fig:lambda}
\end{figure}

Figure \ref{fig:lambda} presents the ordered eigenvalues for the patches, of size $8 \times 8$, of each of the $49$ images in the data set separately as well as for the complete ensemble, \ie, the {\tt aggregate} in black. The aggregate $\lambda$ plot is representative of the individual plots for the small indices $1,2,\ldots,m$. Hence, it is reasonable to use the aggregate value in the design. The images in the data set have more varied individual $\lambda_j$ values as $j$ grows larger but this will not significantly influence the quality of recovery as $\sigma_n^2 > \lambda_j$ by then. Table~\ref{table:lambda} gives the first $8$ eigenvalues, rounded to three decimal places, of six images from Figure~\ref{fig:images} and of {\tt aggregate}.

\begin{table}[h!]
\caption{The First $8$ Eigenvalues when $M=64$.}
\label{table:lambda}
\centering
\begin{tabular}{c l c }
\hline
No. & Image & $\lambda_1 \quad \lambda_2 \quad \ldots \quad \lambda_8$\\
\hline
& {\tt aggregate} & $10.216 \quad 0.244 \quad 0.238 \quad 0.096 \quad 
0.080 \quad 0.078 \quad 0.051 \quad 0.045$\\
		
1 & {\tt merlion} & $15.125 \quad 0.245 \quad 0.142 \quad 0.096 \quad 
0.056 \quad 0.043 \quad 0.039 \quad 0.035$\\
		
2 & {\tt dragon} & $10.590 \quad 0.454 \quad 0.333 \quad 0.200 \quad 
0.164 \quad 0.122 \quad 0.107 \quad 0.078$\\

3 & {\tt lena} & $10.115 \quad 0.078 \quad 0.044 \quad 0.014 \quad 
0.010 \quad 0.005 \quad 0.004 \quad 0.002$\\
		
4 & {\tt flood} & $~~8.714 \quad 0.216 \quad 0.167 \quad 0.047 \quad 
0.042 \quad 0.041 \quad 0.017 \quad 0.016$\\
		
5 & {\tt mandrill} & $~~4.276 \quad 0.326 \quad 0.280 \quad 0.201 \quad 
0.154 \quad 0.138 \quad 0.114 \quad 0.100$\\
		
6 & {\tt oldlady} & $~~2.578 \quad 0.080 \quad 0.043 \quad 0.025 \quad 
0.020 \quad 0.014 \quad 0.014 \quad 0.012$\\
\hline
\end{tabular}
\end{table}	

\begin{figure}[th!]
	\centering
	\begin{tabular}{cc}
		\includegraphics[width=0.45\linewidth]{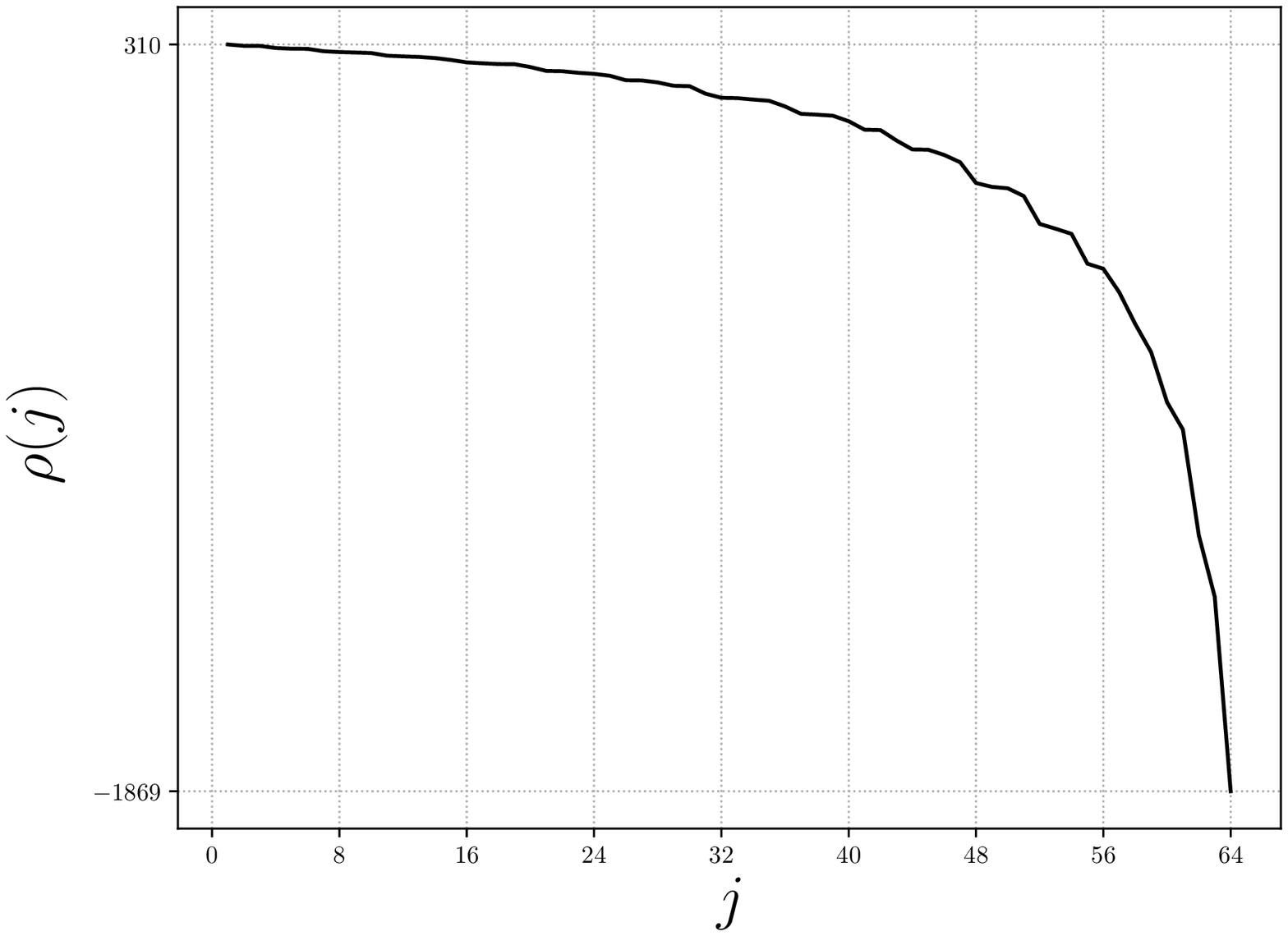} &
		\includegraphics[width=0.45\linewidth]{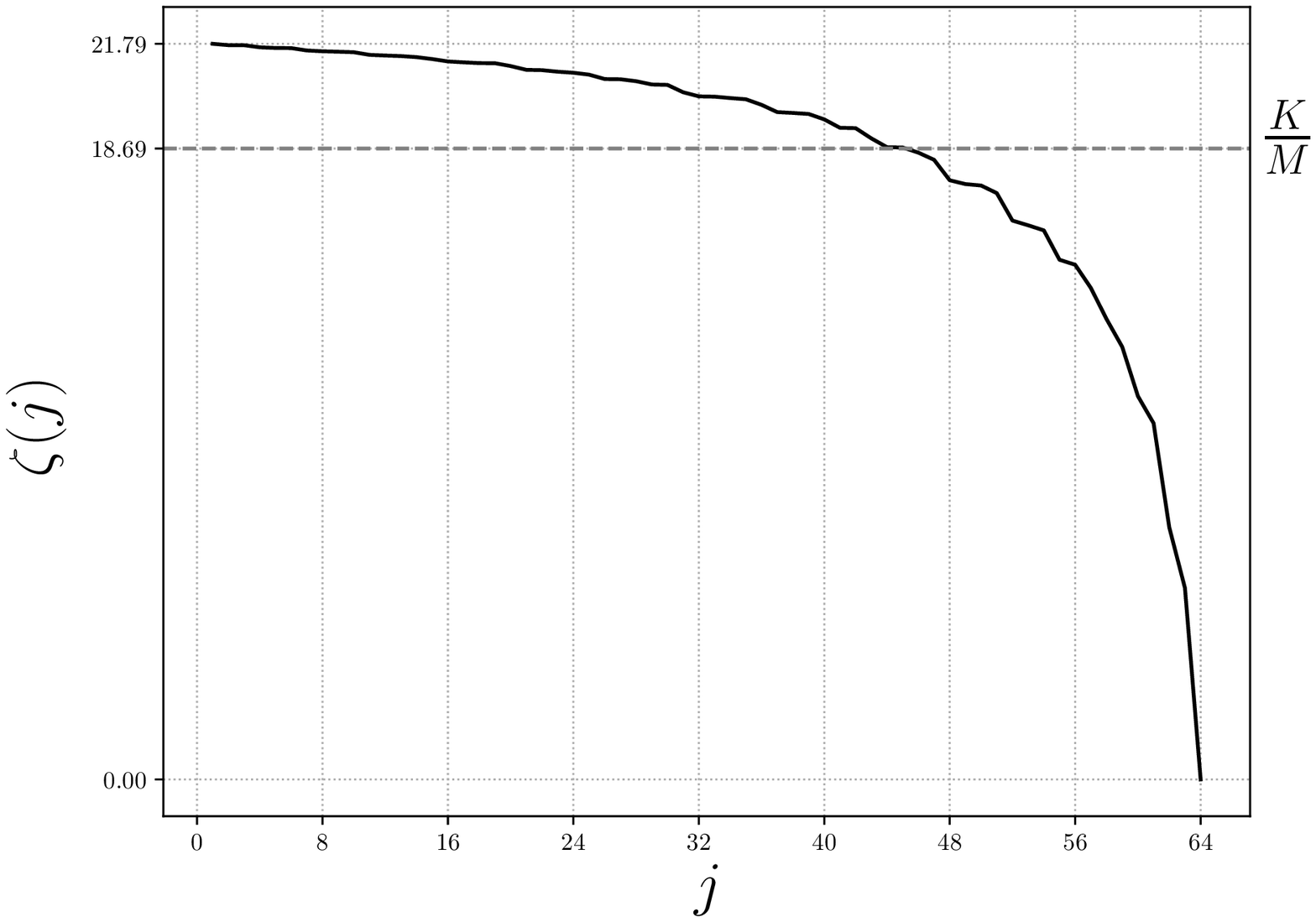}\\ 
		(a) The plot $\rho(j)$ of {\tt aggregate} & (b) $\zeta_j$ when $\sigma_n^2=0.01$\\
		\includegraphics[width=0.45\linewidth]{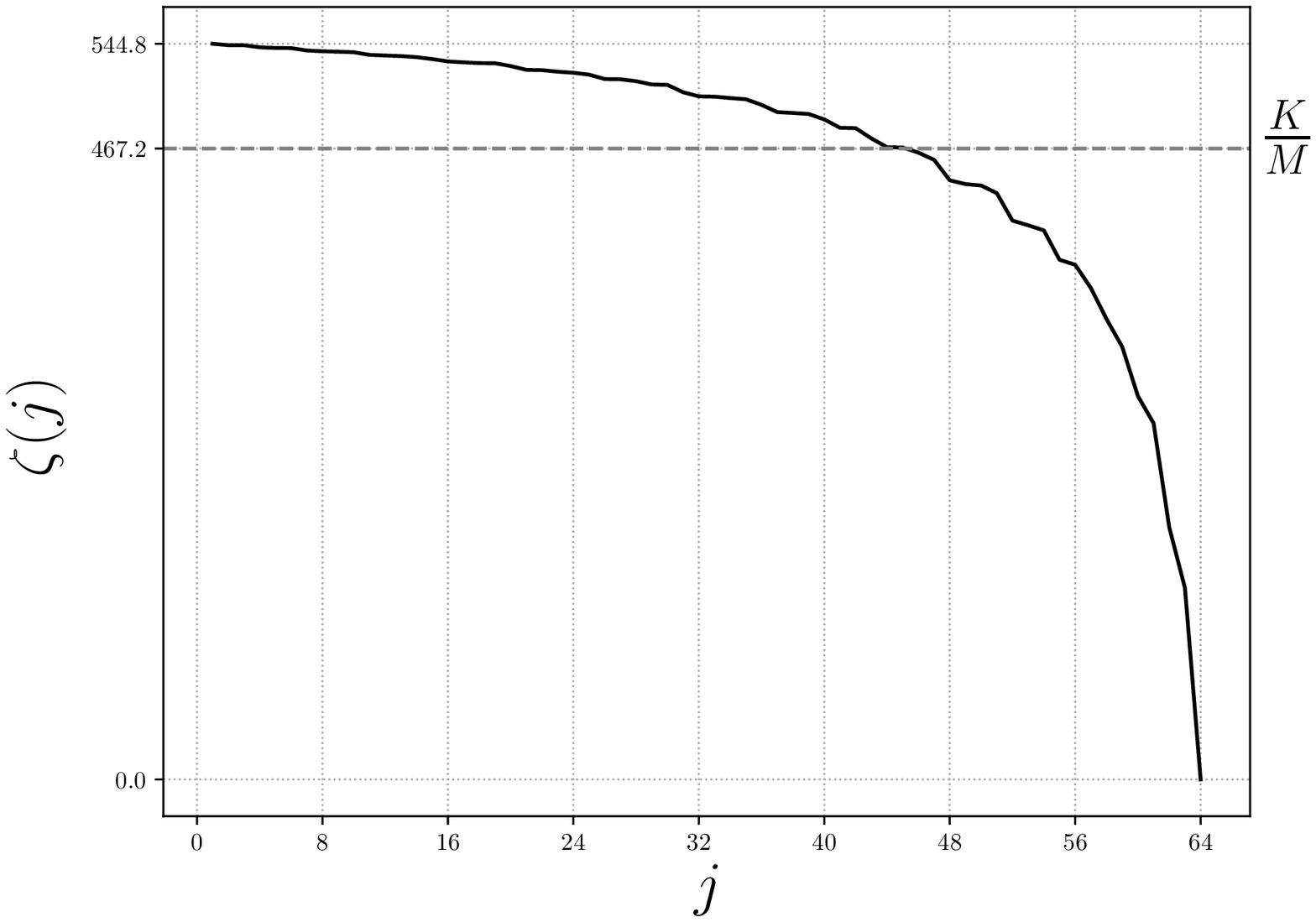} &
		\includegraphics[width=0.45\linewidth]{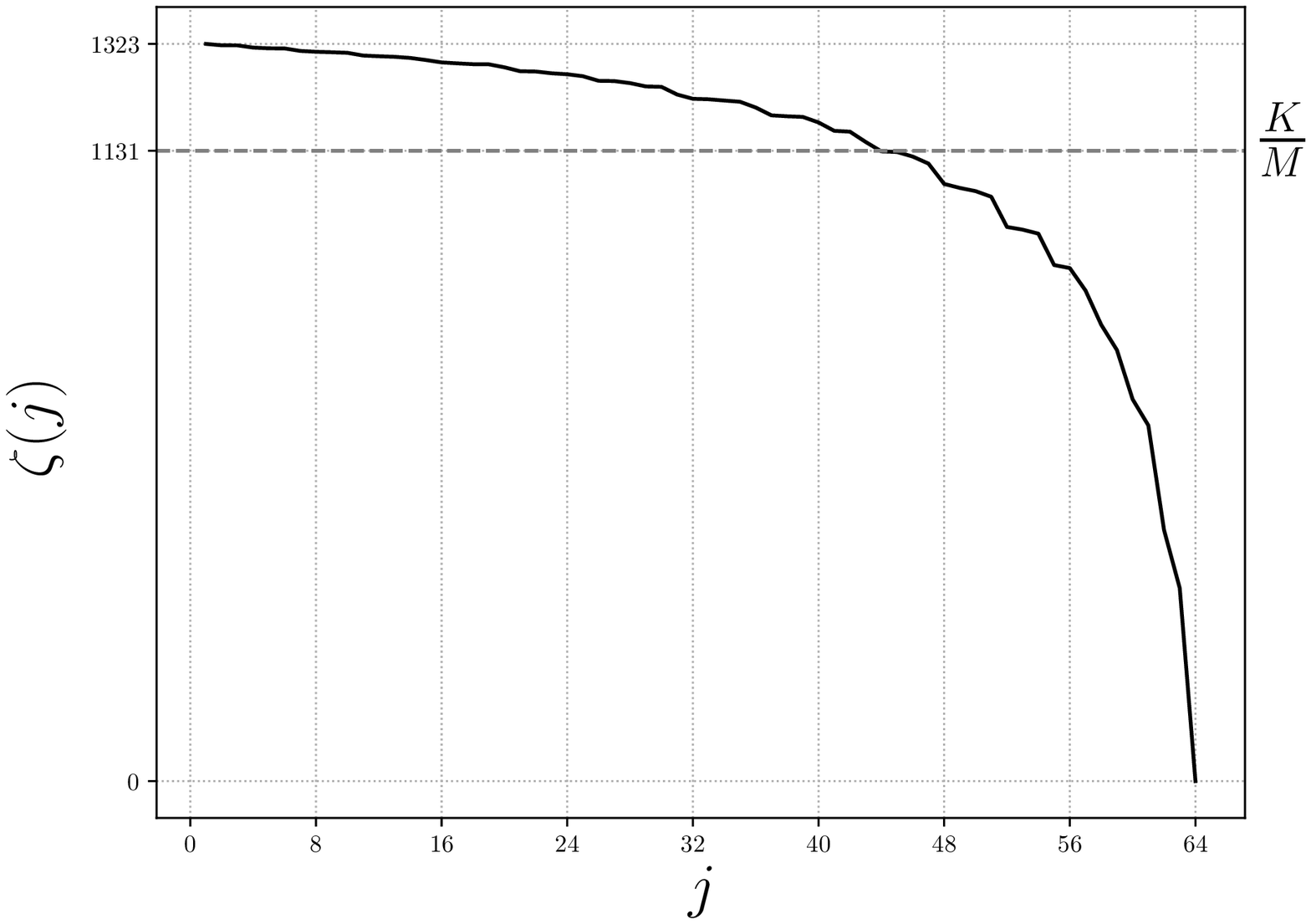}\\ 
		(c) $\zeta_j$ when $\sigma_n^2=0.25$ & (d) $\zeta_j$ when $\sigma_n^2=0.64$\\
	\end{tabular}
	\caption{The plot of the function $\rho(j)$ and the respective distribution of $\zeta_j$ for $j \in \bbra{64}$ based on the $\Lambda$ profile of {\tt aggregate} for the indicated $\sigma_n^2$.}
	\label{fig:rho_zeta}
\end{figure}

The plot $\rho(j)$ for $j \in \bbra{64}$ based on Equation (\ref{eq:rho}) using the $\lambda_j$ values of {\tt aggregate} is in Figure~\ref{fig:rho_zeta} (a). If we are given just enough number $K$ of probings to allocate such that $\zeta_j \geq 0$ for all $j \in \bbra{64}$, as was discussed in Subsection~\ref{subsec:optprobe} above, then the respective plots of $\zeta_j$ for $\sigma_n^2 \in \{0.01,0.25,0.64\}$ are those given in Figure~\ref{fig:rho_zeta} (b)--(d). Each plot shows how the curve is influenced by $\frac{K}{M}$, which serves as the constant level around which each $\zeta_j$ lives, starting above it before decreasing as $j$ goes to $M$. The shape is determined by $\sigma_n^2 \cdot \rho(j)$ since $\zeta_j = \frac{K}{M} + \sigma_n^2 \cdot \rho(j)$. 

\smallskip
\noindent {\it The Distribution of Probings}
\begin{table}[ht!]
	\caption{The assignments $s_1,\ldots, s_{L}$ based on the $\Lambda$ profiles of {\tt aggregate}, {\tt lena}, and {\tt mandrill}. For Mode $3$, the distribution is always $[N]^m$, \ie, the first $m$ indices are each probed $N$ times.}
	\label{table:distro}
	\renewcommand{\arraystretch}{1.05}
	\setlength{\tabcolsep}{0.1cm}
	\footnotesize
	\centering
	\begin{tabular}{c c c lr }
		\hline
		No. & $(M,m,N)$ & $\sigma_n^2$ & \multicolumn{2}{c}{Distribution when using the statistics of {\tt aggregate}} \\
		&& & Mode $1$ & Mode $2$ \\ 
		\hline
		1 & $(64,4,8)$ & $0.01$ & $[2]^4 [1]^{24}$ & $[8][7][1]^{17}$ \\
		2 & & $0.25$ & $[7][6][5][4][3]^2[2][1]^2$ & $[8]^2[5][3]^3[1]^2$ \\
		3 & & $0.64$ & $[8]^3[4][2]^2$ & $[8]^3[5][2][1]$ \\
		4 & & $1.00$ & $[8]^4$ & $[8]^4$ \\
		
		5 & $(64,4,16)$ & $0.01$ & $[3]^2[2]^{23} [1]^{12}$ & $[16][7][2]^{12}[1]^{17}$\\
		6 & & $0.25$ & $[10][9][8][7][6]^2[5][4]^2[3][1]^2$ & $[16][8]^2[6]^3[4][3]^3[1]$ \\
		7 & & $0.64$ & $[15][12]^2[8][7]^2[2][1]$ & $[16][14][12][8][6]^2[2]$ \\
		8 & & $1.00$ & $[16]^2[14][8][5]^2$ & $[16]^2[15][7][5]^2$ \\
		9 & & $2.00$ & $[16]^4$ & $[16]^4$ \\
		
		10 & $(64,8,8)$ & $0.01$ & $[3]^2 [2]^{23} [1]^{12}$ & $[8]^3 [3] [2]^{10} [1]^{17}$ \\
		11 & & $0.25$ & $[8]^5[6][5][4]^2[3][1]^2$ & $[8]^6[6][3]^3[1]$ \\
		12 & & $0.64$ & $[8]^8$ & $[8]^8$ \\
		
		13 & $(64,8,16)$ & $0.01$ & $[4]^{12}[3]^{18} [2]^{10}[1]^6$ & $[16][15][3]^{22}[2]^{12} [1]^7$ \\
		14 & & $0.25$ & $[14][13]^2[11]^3[9][8]^2[7][6][5]^2[4][2][1]$ & $[16]^3[11][10]^2[8]^2[7]^2[5]^2[4][3][2]$ \\
		15 & & $0.64$ & $[16]^6[14][7][6][5]$ & $[16]^6[14][7][6][5]$ \\
		16 & & $1.00$ & $[16]^7[12][3][1]$ & $[16]^7[13][2][1]$ \\
		17 & & $2.00$ & $[16]^8$ & $[16]^8$ \\
		
		\hline
		No. & $(M,m,N)$ & $\sigma_n^2$ & \multicolumn{2}{c}{Distribution for {\tt lena}} \\
		&& & Mode $1$ & Mode $2$ \\ 
		\hline
		1 & $(64,4,8)$ & $0.01$ & $[5]^4 [4] [3]^2 [2]$ & $[8] [5] [4]^3 [3] [2]^2$ \\
		
		2 & & $0.25$ & $[8]^4$ & $[8]^4$ \\
		
		3 & $(64,4,16)$ & $0.01$ & $[8]^4 [7]^2 [6]^2 [3][2][1]$ & $[13] [8]^2 [7]^2 [6] [5]^2 [3][2]$ \\
		
		4 & & $0.25$ & $[16]^4$ & $[16]^4$ \\
		
		5 & $(64,8,8)$ & $0.01$ & $[8]^4 [7]^2 [6]^2 [3][2][1]$ & $[8]^6 [6] [5] [3][2]$ \\
		
		6 & & $0.25$ & $[8]^8$ & $[8]^8$ \\
		
		7 & $(64,8,16)$ & $0.01$ & $[13]^4 [12][11]^3 [8] [7][5]^2[3][2][1]$ & $[16]^2 [13][12]^2 [11] [10]^2 [7]^2 [5]^2 [2][1]^2$ \\
		
		8 & & $0.25$ & $[16]^8$ & $[16]^8$ \\
		
		\hline
		No. & $(M,m,N)$ & $\sigma_n^2$ & \multicolumn{2}{c}{Distribution for {\tt mandrill}} \\
		&& & Mode $1$ & Mode $2$ \\ 
		\hline
		1 & $(64,4,8)$ & $0.01$ & $[1]^{32}$ & $[8]^3 [7] [1]$\\
		
		2 & & $0.25$ & $[4]^3 [3]^3 [2]^3 [1]^5$ & $[8] [7] [3]^2 [2]^3 [1]^5$\\
		
		3 & & $0.64$ & $[7][6][5][4][3]^2[2][1]^2$ & $[8]^2[5][4][3][2][1]^2$\\
		
		4 & & $1.00$ & $[8][7][6][5][3][2][1]$ & $[8]^3 [4][2]^2$\\
		
		5 & & $2.00$ & $[8]^3[7][1]$ & $[8]^4$\\
		
		6 & $(64,4,16)$ & $0.01$ & $[2]^{15} [1]^{34}$ & $[16] [11][1]^{37}$\\
		
		7 & & $0.25$ & $[6]^2 [5]^4 [4]^3 [3]^4 [2]^2 [1]^4$ & $[15] [5]^3 [4]^3 [3]^4 [2]^3 [1]^4$\\
		
		8 & & $0.64$ & $[10][8]^2[7][6]^2[5][4][3]^2[2][1]^2$ & $[16][8][7]^2[6][5][4][3]^2[2]^2[1]$\\
		
		9 & & $1.00$ & 
		$[13][10][9][8][7][6][4][3][2][1]^2$ & $[16][10][9][8][6][5][4][3][2][1]$\\
		
		10 & & $2.00$ & $[16]^2[12][9][6][4][1]$ & $[16]^2[13][9][5][4][1]$\\
		
		11 & & $4.00$ & $[16]^3[15][1]$ & $[16]^3[15][1]$\\
		
		12 & $(64,8,8)$ & $0.01$ & $[2]^{15} [1]^{34}$ & $[8]^3 [5][1]^{35}$\\
		
		13 & & $0.25$ & $[6]^2 [5]^4 [4]^3 [3]^4 [2]^2 [1]^4$ & $[8]^3 [6] [4]^3 [3]^4 [2]^3 [1]^4$\\
		
		14 & & $0.64$ & $[8]^4[7][6][5][4][3]^2[2][1]^2$ & $[8]^6[5][3]^2[2]^2[1]$\\
		
		15 & & $1.00$ & $[8]^7[4][2][1]^2$ & $[8]^7[5][2][1]$\\
		
		16 & & $2.00$ & $[8]^8$ & $[8]^8$\\
		
		17 & $(64,8,16)$ & $0.01$ & $[3]^{19} [2]^{31}[1]^9$ & $[16]^2 [5][2]^{37}[1]^{17}$\\
		
		18 & & $0.25$ & $[9][8]^4 [7]^3 [6]^4 [5]^2 [4]^5 [3]^2[2]^2[1]^2$ & $[16] [12][8][7]^3 [6]^5 [5]^3 [4]^3 [3]^2[2]^3[1]^2$\\
		
		19 & & $0.64$ & $[14][12]^2[11][10]^2[9][8]^2[7][6]^2[5][4][2]^2[1]^2$ & $[16]^3[11][10][9][8][7]^2[6]^2[5][4]^2[1]^3$\\
		
		20 & & $1.00$ & $[16]^2[15][13][12][11][9][8][7][6]^2[4][3][2]$ & $[16]^4[11][10][9][7]^2[6][5][4][3][2]$\\
		
		21 & & $2.00$ & $[16]^7[7][5][3][1]$ & $[16]^7[9][4][2][1]$\\
		
		22 & & $4.00$ & $[16]^8$ & $[16]^8$\\
		
		\hline
		
	\end{tabular}
\end{table}

For the various levels of noise we next determine the projection packet designs to holographically sense the images in the data set, based on the learned second order statistics $\BR_{xx}$. Since we aim to achieve the lowest $\MSE$ when all $N$ packets, \ie, $K=N \cdot m$ probings, are available, we solve for $\zeta_j$ for $j \in \bbra{L}$. Calibrations to conform to Modes $1$ to $3$ are subsequently carried out.

Table~\ref{table:distro} presents the optimal sensing distributions obtained for the indicated noise levels, based on either the $\Lambda$ of {\tt aggregate}, or the $\Lambda$ of {\tt lena}, or that of {\tt mandrill}, for illustrative comparison. The set $\{s_j : j \in \bbra{L}\}$ is written in shorthand with $[x_1]^{y_1}[x_2]^{y_2} \ldots [x_r]^{y_r}$ denoting $s_j=x_1$ for $j \in \bbra{y_1}$ followed by $s_j=x_2$ for $j \in \bbra{y_1+1,y_1+y_2}$ and so on until $s_j=x_r$ for $j \in \bbra{L-y_r+1,L}$. We remove the superscript if $y_i=1$. Notice that $L = \sum_{i=1}^r y_r$. For example, Entry 3 in Table~\ref{table:distro} for the {\tt aggregate}'s statistics has the distribution $[8]^3[4][2]^2$ in Mode $1$. We read this as $L=6$ with $s_1=s_2=s_3=8$, $s_4=4$, and $s_5=s_6=2$, when all $N$ packets are available. 

\begin{example}
Let $(M,m,N)=(64,8,8)$. Using Mode $1$ on the $\Lambda$ profile of {\tt aggregate}, we obtain the following packet sensing allocations for $\sigma_n^2 \in \{0.01, 0.25, 0.64\}$, where locations refer to the indices $i$ where $\BP_{k}$ has entry $1$ on its diagonal. 
	
\smallskip
\centering{
\begin{tabular}{cccc}
$k$ & Locations for $\sigma_n^2=0.01$  & Locations for $\sigma_n^2=0.25$ & Locations for $\sigma_n^2=0.64$ \\
\hline
$1$ & $1, 4, 8, 12, 16, 20, 24, 30$ & $1, 2, 3, 4, 5, 6, 7, 9$ & $1, 2, 3, 4, 5, 6, 7, 8$ \\
$2$ & $1, 4, 8, 12, 16, 20, 24, 31$ & $1, 2, 3, 4, 5, 6, 7, 9$ & $1, 2, 3, 4, 5, 6, 7, 8$ \\
$3$ & $1, 5, 9, 13, 17, 21, 25, 32$ & $1, 2, 3, 4, 5, 6, 7, 9$ & $1, 2, 3, 4, 5, 6, 7, 8$ \\
$4$ & $2, 5, 9, 13, 17, 21, 25, 33$ & $1, 2, 3, 4, 5, 6, 8, 10$ & $1, 2, 3, 4, 5, 6, 7, 8$ \\
$5$ & $2, 6, 10, 14, 18, 22, 26, 34$ & $1, 2, 3, 4, 5, 6, 8, 10$ & $1, 2, 3, 4, 5, 6, 7, 8$ \\
$6$ & $2, 6, 10, 14, 18, 22, 27, 35$ & $1, 2, 3, 4, 5, 6, 8, 10$ & $1, 2, 3, 4, 5, 6, 7, 8$ \\
$7$ & $3, 7, 11, 15, 19, 23, 28, 36$ & $1, 2, 3, 4, 5, 7, 8, 11$ & $1, 2, 3, 4, 5, 6, 7, 8$ \\
$8$ & $3, 7, 11, 15, 19, 23, 29, 37$ & $1, 2, 3, 4, 5, 7, 9, 12$ & $1, 2, 3, 4, 5, 6, 7, 8$ \\
\end{tabular}}
\end{example}

\subsection{The Sensing and Recovery Steps}
Now that all of the ingredients to sense any input image are in place, we are ready to showcase the holographic image sensing in action. Several functionalities are included in the software for analytical purposes. Aside from allowing users to vary the basic parameters $M$, $m$, $N$, and $\sigma_n^2$, there are options to perform the sensing and complete the recovery process by using either the $\Lambda$ of {\tt aggregate} or that of the individual image. The modes can also be set as desired. Recall that we represent $\by_{\omega}$, instead of $\bx_{\omega}$, holographically, as $N$ packets of data $\bz_k$ given by
\[
\Big\{ \bz_{k} = \BU_{k} \Psi^{\top} \bx_{\omega} + \bn_{k} : k \in \bbra{N} \Big\}.
\]
The packets can be stored for later recovery, once needed, or transmitted over some channels for reconstruction at another location.

In the recovery process, we assume that an arbitrary $\ell$ out of $N$ packets have been made available. We have $\BU_{\mbox{combi},\kappa} := \left(\BU_{k_1}| \BU_{k_2}|\ldots|\BU_{k_{\ell}}\right)$, where $\kappa:=\{k_1,k_2,\ldots,k_{\ell}\}$ is the corresponding index set. The recovered estimate $\widehat{\by}_{\omega}$ of $\by_{\omega}$ is given by
\begin{equation}
\widehat{\by}_{\omega,\kappa} = \Lambda ~\BU_{\mbox{combi},\kappa}~ \BM^{-1} ~\bz_{\mbox{combi},\kappa} \mbox{ where }
\BM := \BU_{\mbox{combi},\kappa}^{\top}~\Lambda ~\BU_{\mbox{combi},\kappa} + \sigma_n^2 \BI_{(\ell \cdot m)}.
\end{equation}
The original vector $\bx_{\omega}$ is thus estimated by $\widehat{\bx}_{\omega,K} := \Psi \widehat{\by}_{\omega,K}$. To recover the whole image we assume that the location of each patch is known, \ie, the index set $\{\omega\}$ is known and in order. Reversing the prosess described at the start of Subsection~\ref{subsec:prep} yields the recovered image. We make available two procedures. 
\begin{enumerate}[wide, itemsep=0pt, leftmargin =0pt,widest={{\bf Randomized}}]
\item[{\bf Incremental}:] Begin by randomly choosing any $1$ out of the $N$ packets and perform the image reconstruction. At each $\ell$, as $\ell$ goes from $2$ to $N$, choose $1$ packet at random from among the remaining $N - (\ell-1)$ packets. Combine the packet with the existing $\bz_{\mbox{combi}}$ and use the result to output an improved image. In this build up, once a packet had been chosen it remains in use for an image reconstruction as $\ell$ increases.
\item[{\bf Randomized}:] As $\ell$ goes from $1$ to $N$, randomly choose any one of the $\binom{N}{\ell}$ possible index sets $\kappa$ of size $\ell$. Use the corresponding $\bz_{\mbox{combi},\kappa}$ in the image reconstruction. Here, a packet which has been included earlier may be dropped in the next iteration. 
\end{enumerate}

\begin{figure}[h!]
\centering
\begin{tabular}{ccc}
	\includegraphics[width=0.3\linewidth]{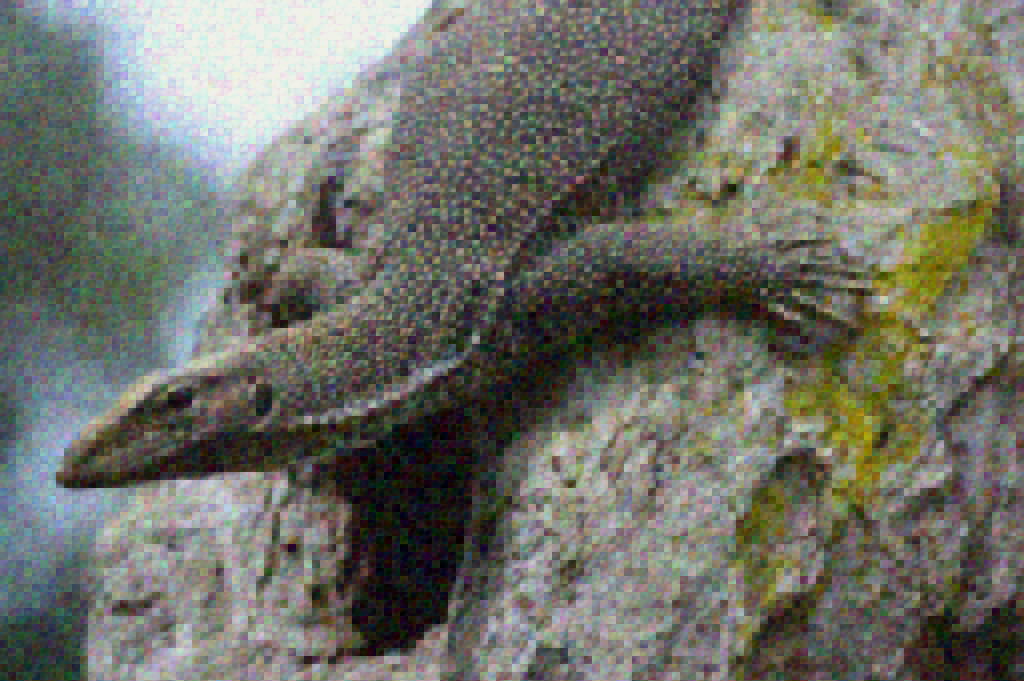} & 
	\includegraphics[width=0.3\linewidth]{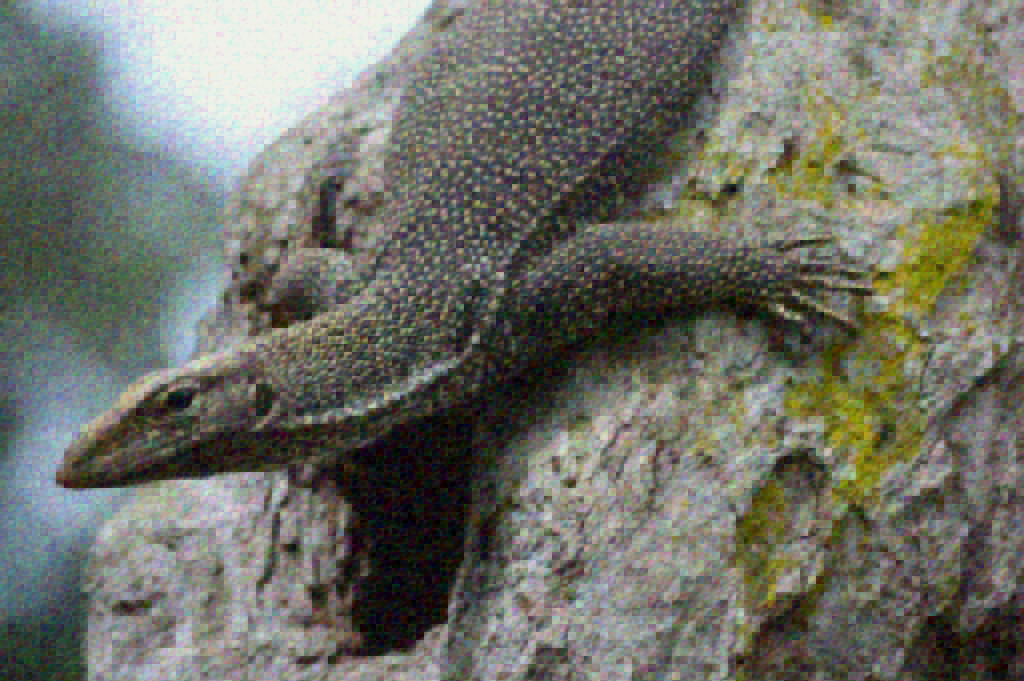} &
	\includegraphics[width=0.3\linewidth]{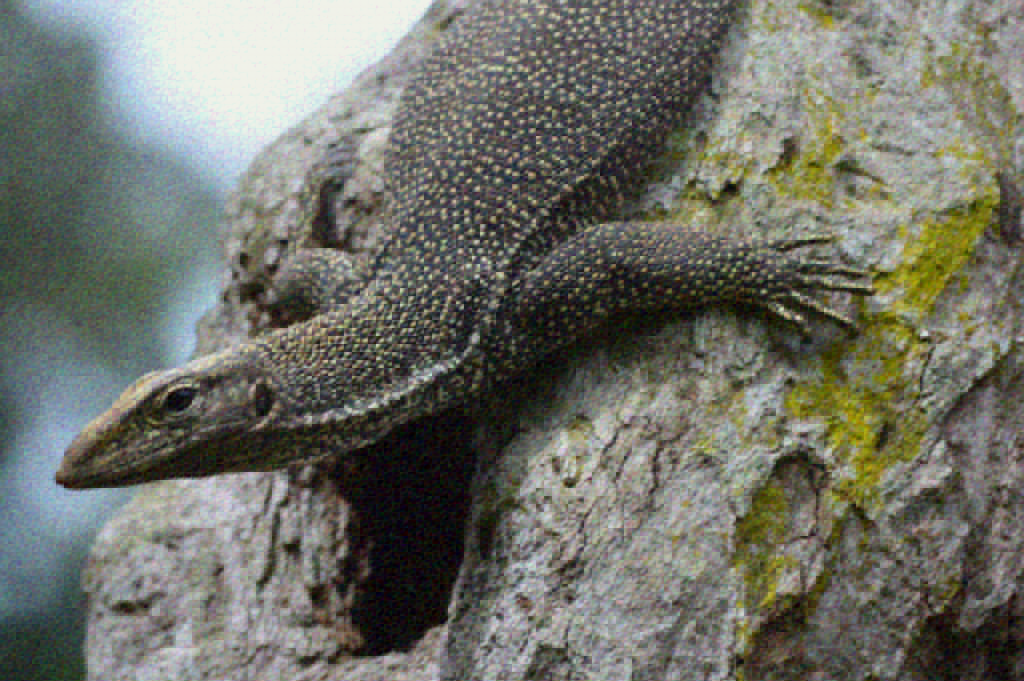} \\
	(a) $\ell=1$ & (b) $\ell=2$ & (c) $\ell=8$: all\\
	
	\includegraphics[width=0.3\linewidth]{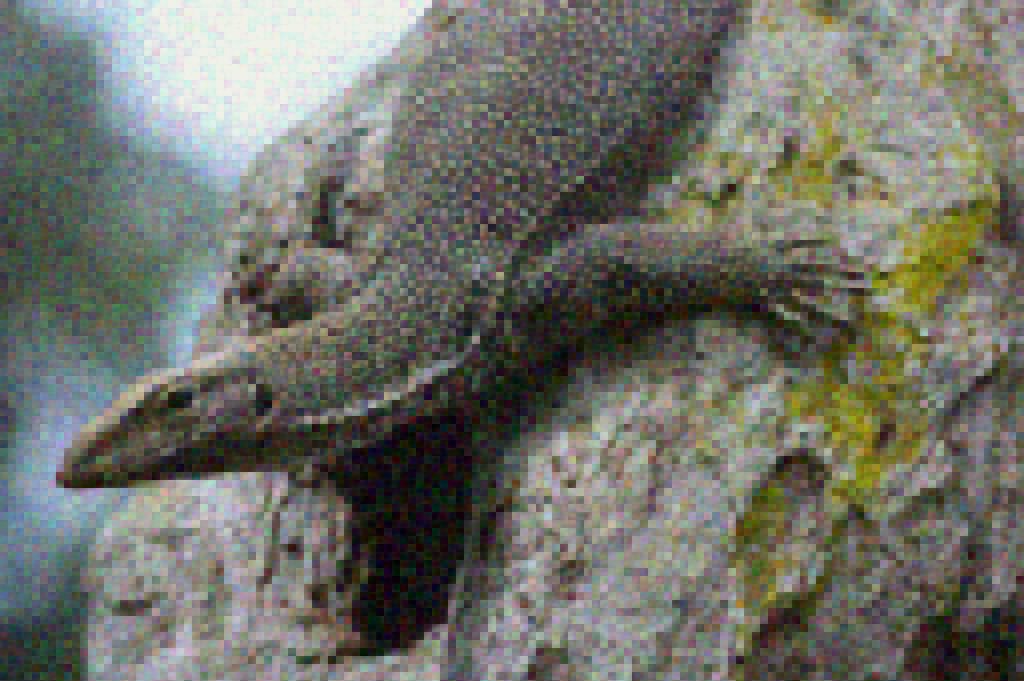} & 
	\includegraphics[width=0.3\linewidth]{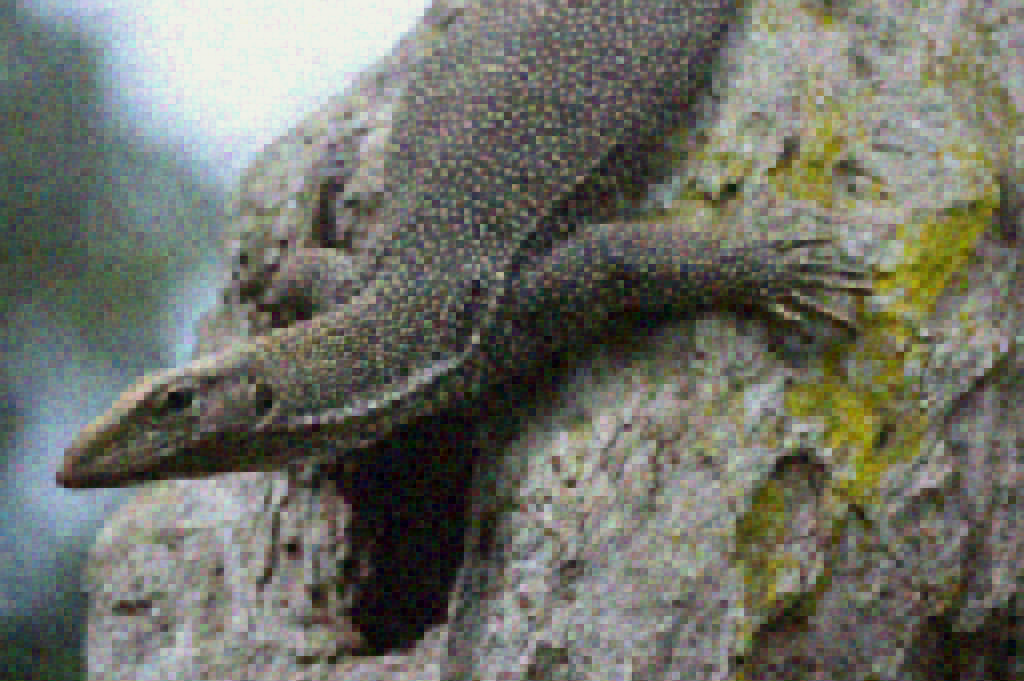} &
	\includegraphics[width=0.3\linewidth]{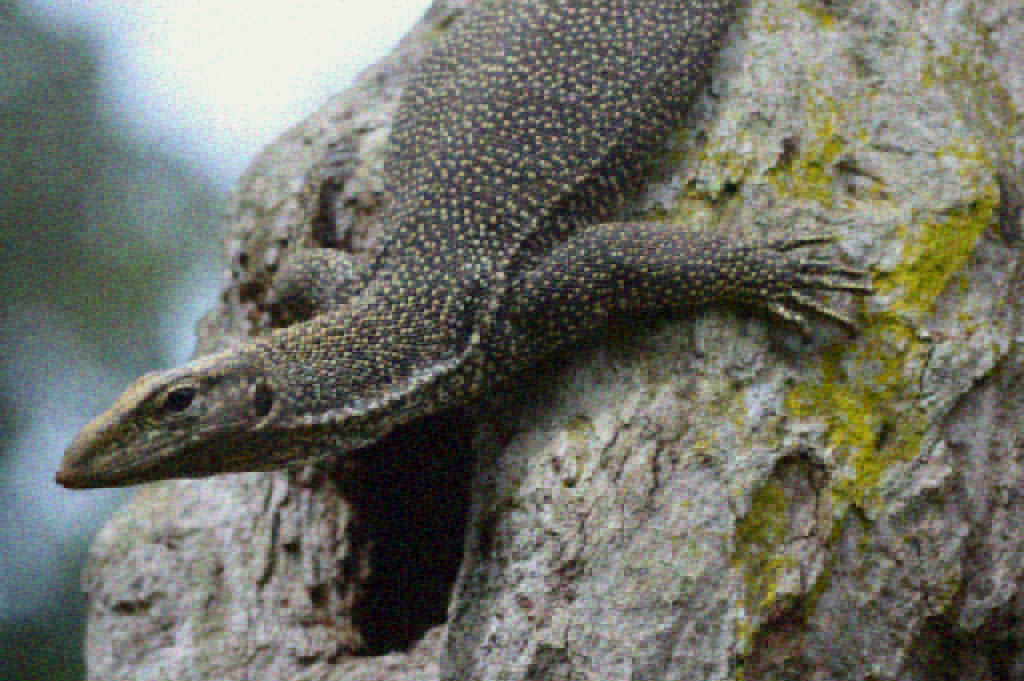} \\
	(d) $\ell=1$ & (e) $\ell=2$ & (f) $\ell=8$: all \\
\end{tabular}
\caption{Recovered {\tt dragon} in the randomized procedure with $(M,m,N,\sigma_n^2)=(64,8,8,0.64)$  on Mode $1$. Recovery for the images in the upper row uses the $\Lambda$ of {\tt dragon}, with distribution $[8]^7[5][2][1]$, while for those in the lower row uses the $\Lambda$ of {\tt aggregate}, with distribution $[8]^8$.}
\label{fig:recovered}
\end{figure}

\begin{figure}[t!]
\centering
\begin{tabular}{ccc}
\includegraphics[width=0.3\linewidth]{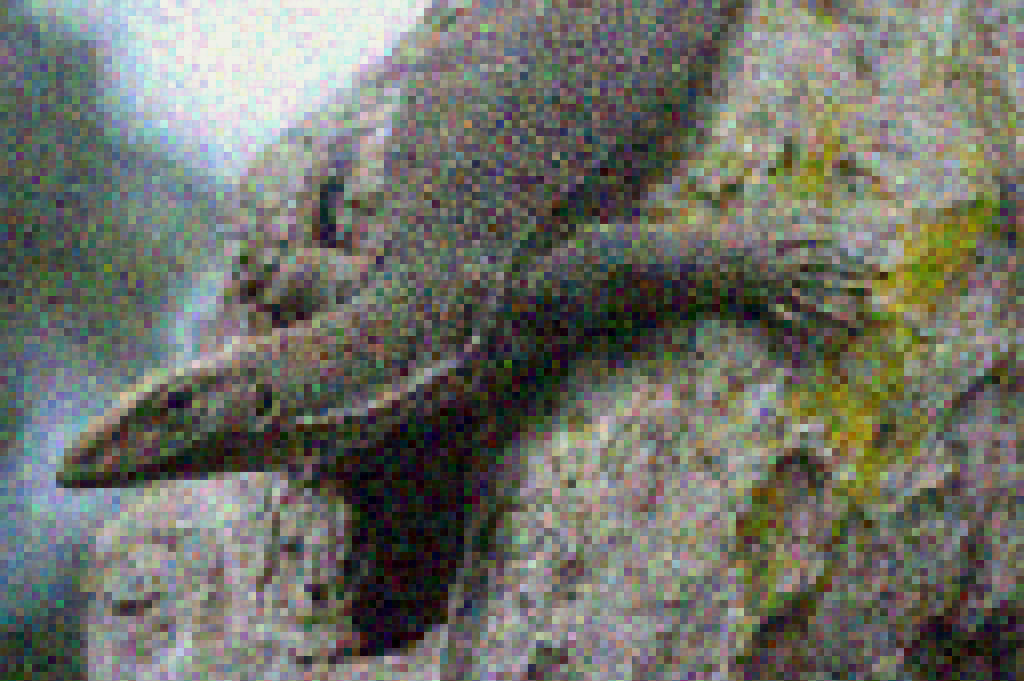} & 
\includegraphics[width=0.3\linewidth]{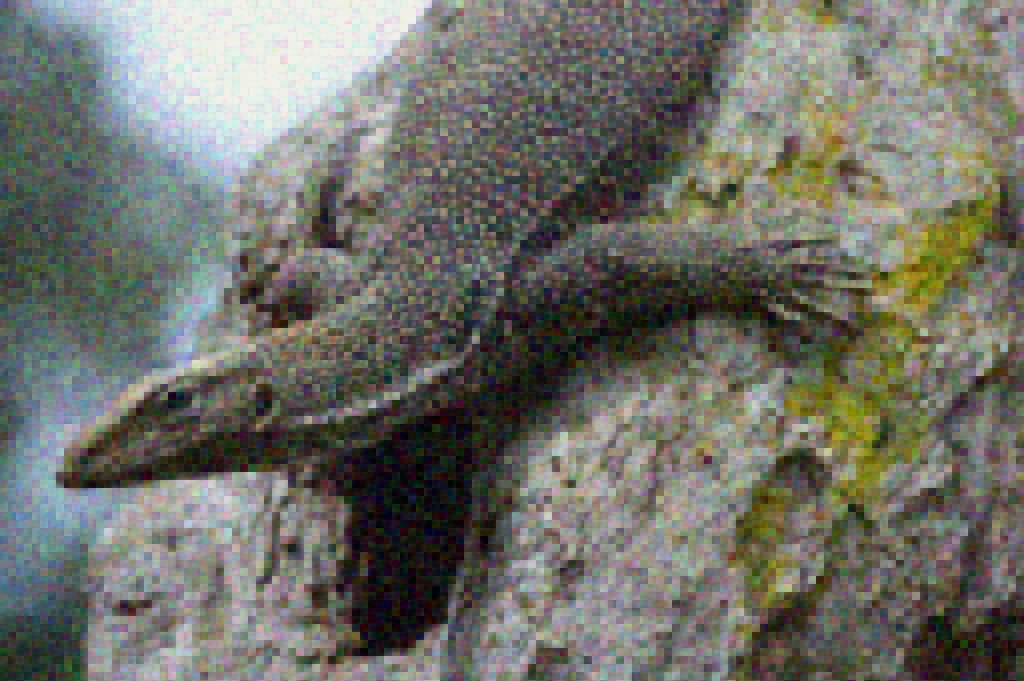} &
\includegraphics[width=0.3\linewidth]{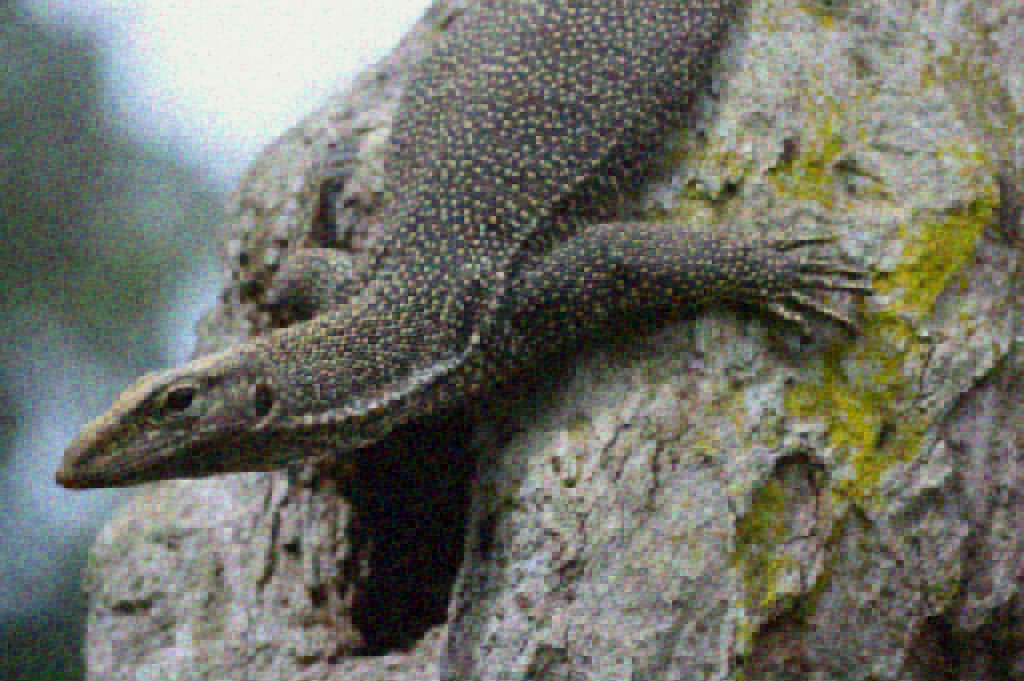} \\
(a) $\ell=1$ & (b) $\ell=2$ & (d) $\ell=8$: all \\
\end{tabular}
\caption{Incrementally recovered {\tt dragon} using the $\Lambda$ of {\tt aggregate} with $(M,m,N,\sigma_n^2)=(64,8,8,1.00)$. The noise forces all three modes to use the distribution $[8]^8$, \ie, $s_j=8$ for $j \in \bbra{8}$ and $s_j=0$ for $j >8$.}
\label{fig:noise1}
\end{figure}

Figure~\ref{fig:recovered} presents recovered {\tt dragon} images using a run of the randomized procedure on Mode $1$ for the supplied input parameters. For a fixed $\ell$, we can see how the resulting images are very similar on recovery based on the two $\Lambda$ profiles, one belonging to {\tt dragon} itself while the other is that of {\tt aggregate}. As the noise level rises, $L$ tends to decrease, favouring the probings of the first few coordinates where the corresponding $\lambda$ values are higher. When $(M,m,N, \sigma_n^2)=(64,8,8,1.00)$, for example, all three modes based on the $\Lambda$ profile of {\tt aggregate} coincide, with sensing distribution $[8]^8$. Figure~\ref{fig:noise1} presents incrementally recovered {\tt dragon} images, on the specified input parameters, for $\ell \in \{1,2,8\}$.

\subsection{Performance Analysis}
To analyze the recovery performance, one can consider several types of $\MSE$ computations in each of the procedures. Let the parameters $M$, $m$, $N$, and $\sigma_n^2$ as well as the mode be given for a chosen image. Hence, we have the $\Lambda$ and $\Psi$ matrices as well as the probing distribution. First, there is the theoretical best $\MSE_{\mbox{best}, \ell}$ from Equation (\ref{eq:best_ell}) to be used as a benchmark. Based on this equation, for a fixed mode, one can compute for the best case $\MSE_{\ell}$ by replacing each $\zeta_j$ by the corresponding $s_j$. Second, for each $\ell$, one can compute the expected value
\begin{equation}\label{eq:expectedMSE}
\EE(\MSE,\ell) := \frac{1}{\binom{N}{\ell}} \sum_{i=1}^{\binom{N}{\ell}}
\left(\frac{1}{\abs{\Omega}}\sum_{\omega=1}^{\abs{\Omega}} 
\left(\bx_{\omega}-\widehat{\bx}_{\omega,i}\right)^2\right)
\end{equation}
with
\[
\left(\bx_{\omega}-\widehat{\bx}_{\omega,i}\right)^2 =
\left(\bx_{\omega}-\widehat{\bx}_{\omega,i}\right)^{\top} \left(\bx_{\omega}-\widehat{\bx}_{\omega,i}\right).
\]
The second index $i$ of $\widehat{\bx}_{\omega,i}$ refers to the $i\textsuperscript{th}$ recovery simulation. For each $\ell$ we compute the average $\MSE$, taken over 
$\min\left(100,\binom{N}{\ell}\right)$ simulations, since averaging over $\binom{N}{\ell}$ simulations may not be practical as $N$ grows.

\begin{figure}[h!]
\centering
\begin{tabular}{cc}
	\includegraphics[width=0.47\linewidth]{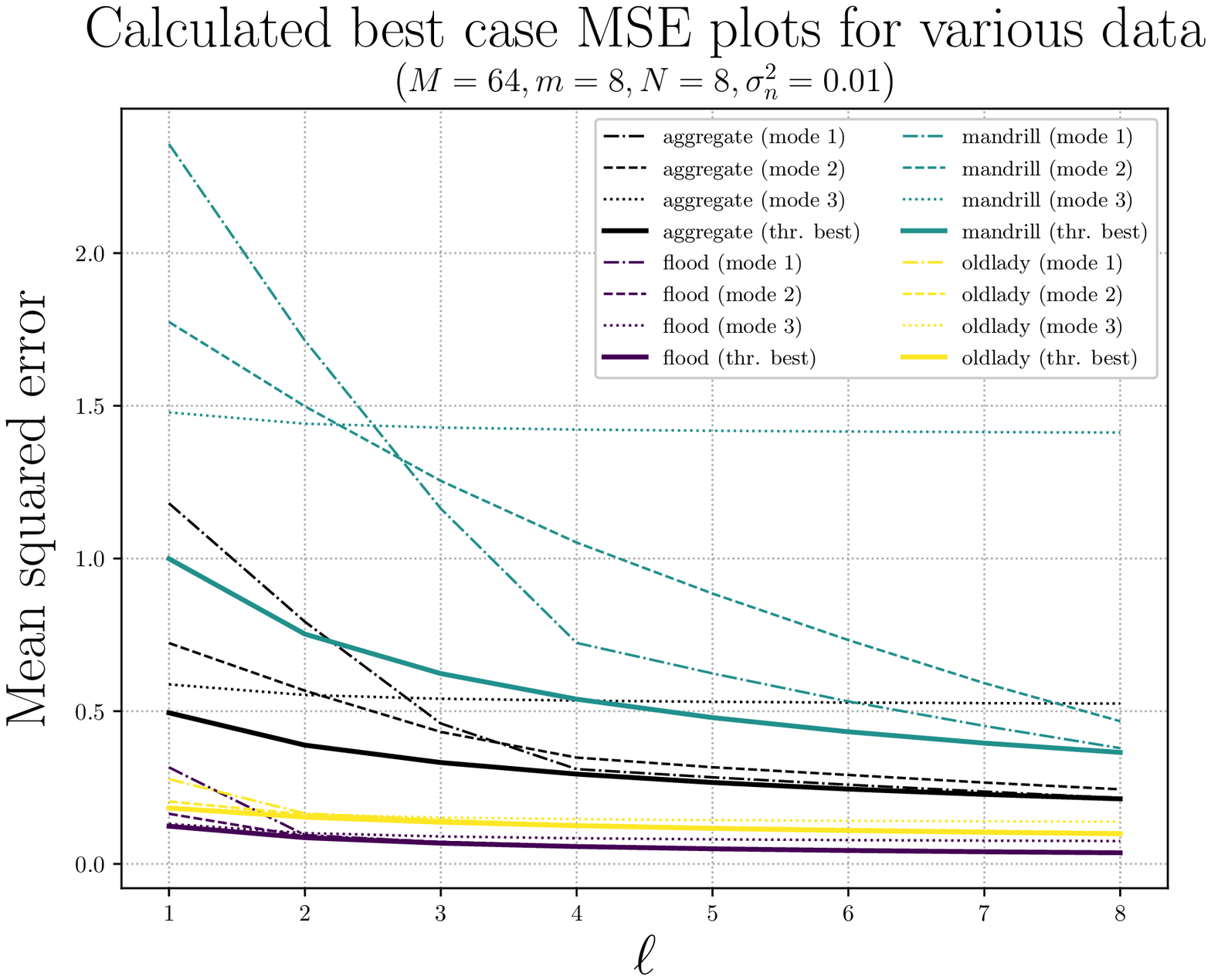} & 
	\includegraphics[width=0.47\linewidth]{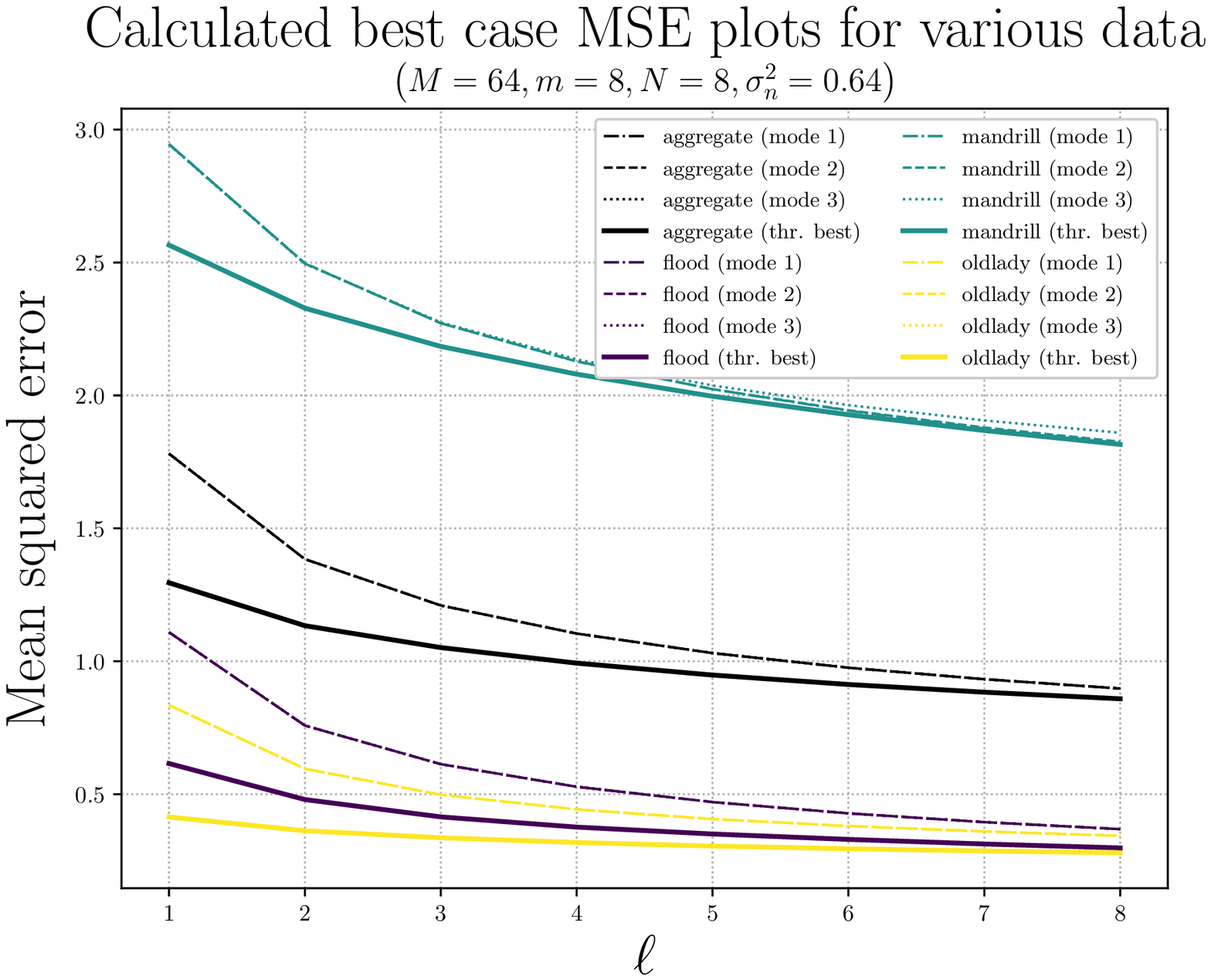} \\
	(a) $\MSE$ when $\sigma_n^2=0.01$ & 
	(b) $\MSE$ when $\sigma_n^2=0.64$ 
\end{tabular}
\caption{A comparison of the plots for the theoretical best (thr. best), computed based on $\zeta_j$ for $j \in \bbra{M}$, for $\ell \in \bbra{N}$, in Equation (\ref{eq:best_ell}), with the computed best case $\MSE$ in the three modes, when $\zeta_j$ is replaced by the respective $s_j$. We use $(M,m,N)=(64,8,8)$ and $\sigma_n^2 \in \{ 0.01, 0.64 \}$.}
\label{fig:comparemode}
\end{figure}

Figure~\ref{fig:comparemode} shows how the best case $\MSE$ curves of the three modes compare with respect to the theoretical best. Mode $1$ comes closest to the theoretical ideal at $\ell=N$, as per the design. When $\ell$ is very small, however, its best case $\MSE$ is considerably higher than those of the other two modes. Mode $2$ has been designed to improve the performance for such $\ell$ at the relatively small cost of decreased performance in the second half of $\ell$ values compared with Mode $1$. Mode $3$ has the best performance at $\ell=1$ but then does not improve as much as the other two modes as more packets become available. 

\begin{figure}[h!]
\centering
\begin{tabular}{cc}
\includegraphics[width=0.47\linewidth]{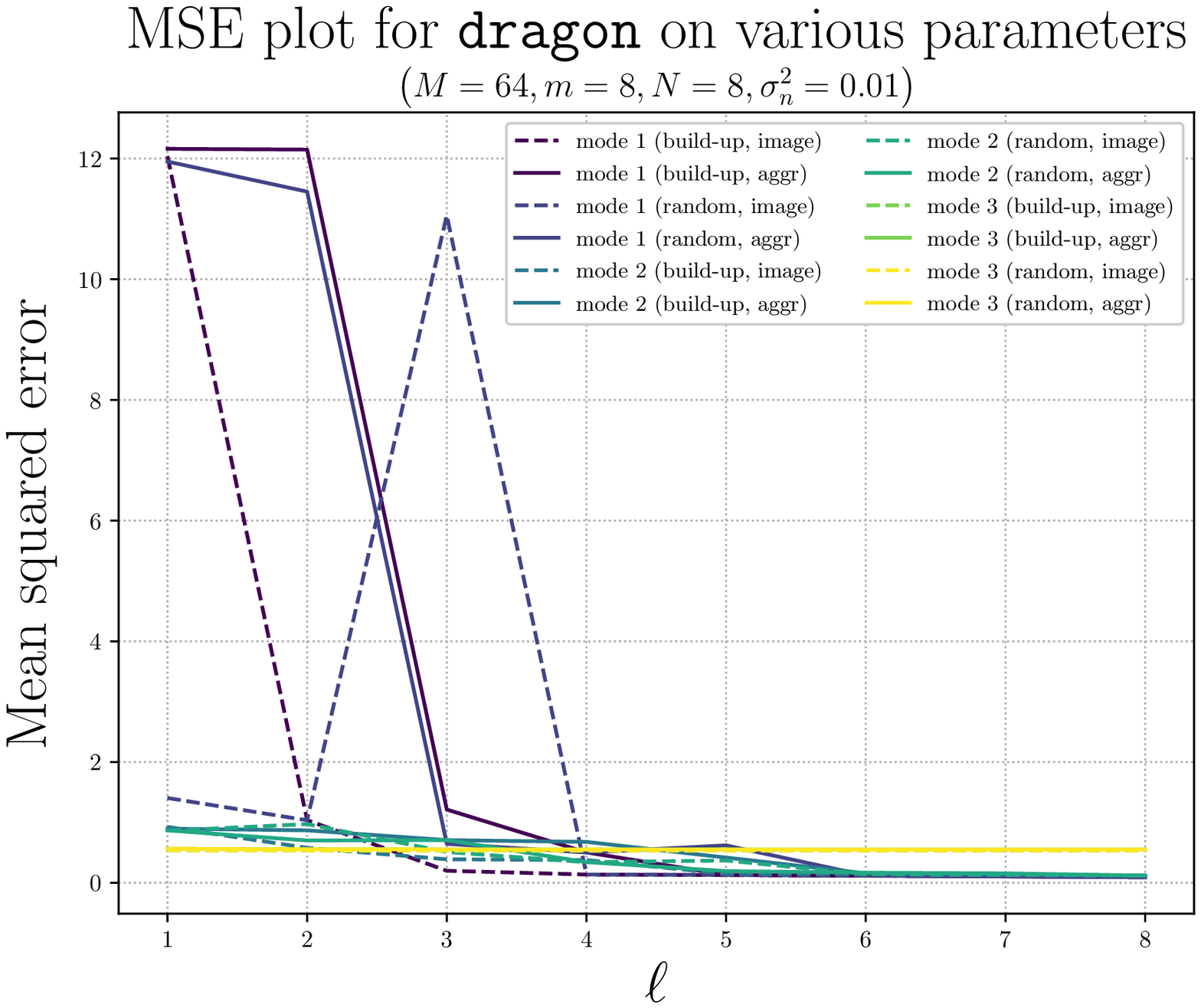} & 
\includegraphics[width=0.47\linewidth]{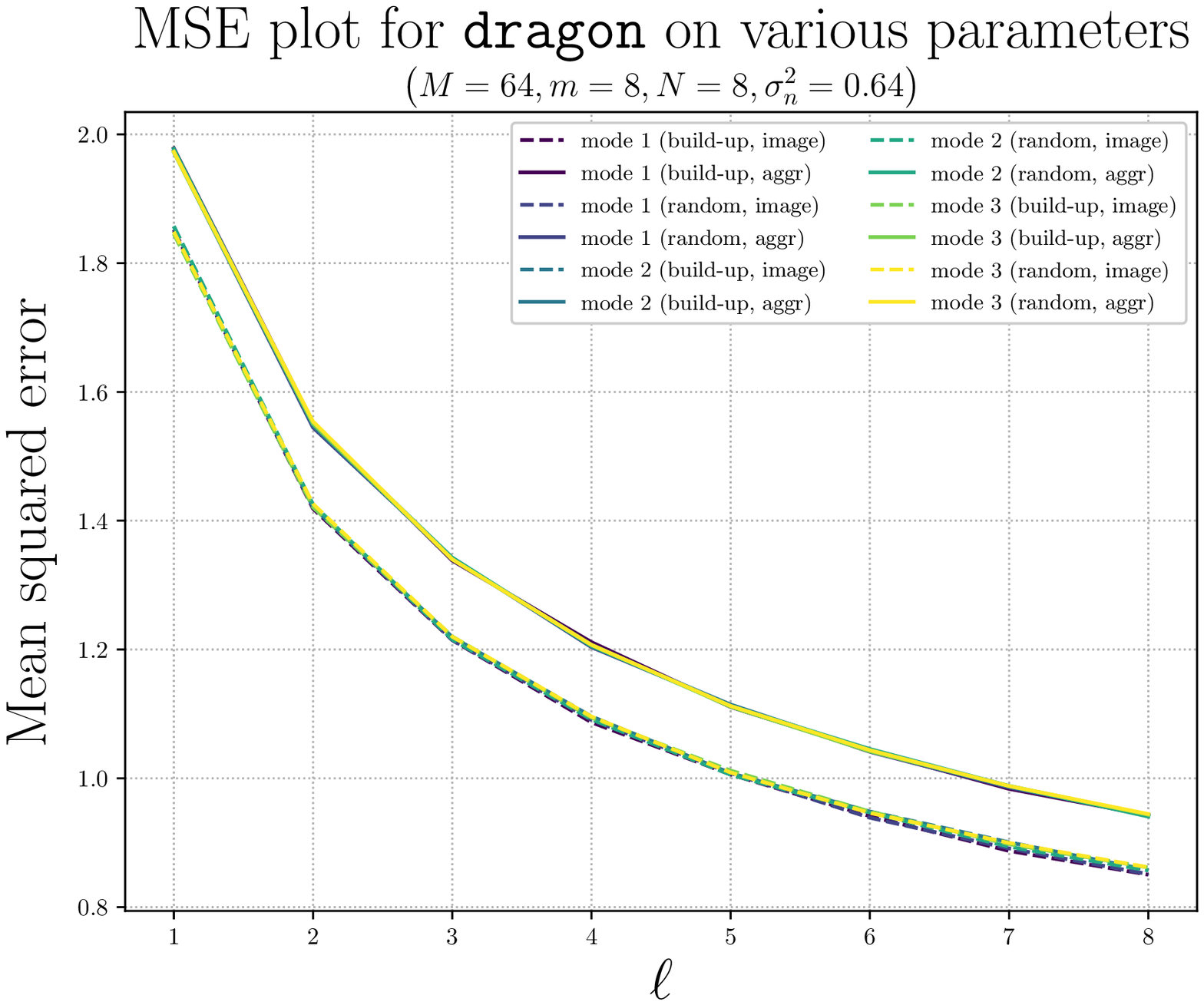} \\
(a) A {\tt dragon} recovery at $\sigma_n^2=0.01$ & 
(b) A {\tt dragon} recovery at $\sigma_n^2=0.64$\\
\end{tabular}
\caption{The trend from a single instance of recovery for each of the indicated setting choices on {\tt dragon} for $(M,m,N)=(64,8,8)$ and $\sigma_n^2 \in \{0.01, 0.64\}$.}
\label{fig:single}
\end{figure}

As we switch from Mode $1$ to $2$ and then to $3$, the $\MSE$ variances among the simulations for small $\ell$ decreases, \ie, the recovery is smoother. There is less chance of recovering the image badly in the early stages of any single simulation. On the other hand, the gain in $\MSE$ reduction as $\ell$ goes toward $N$ also lessens. Higher noise forces the plots of the three modes closer together, often to the point of merging into one plot. This is because the sensing mechanism assigns more probings to the portion of $\by_{\omega}$ with larger covariances to mitigate the effect of higher noise. 

For an easier inspection on how a change in noise level affects the performance in an actual recovery, Figure~\ref{fig:single} illustrates the trend on {\tt dragon} for $(M,m,N)=(64,8,8)$ and $\sigma_n^2 \in \{0.01, 0.64\}$. For each mode, we run a single recovery procedure based on the $\Lambda$ of {\tt dragon} and the $\Lambda$ of {\tt aggregate}. Relevant plots for other images and parameter combinations can be similarly generated.

\begin{figure}[h!]
\centering
\begin{tabular}{cc}	
	\includegraphics[width=0.47\linewidth]{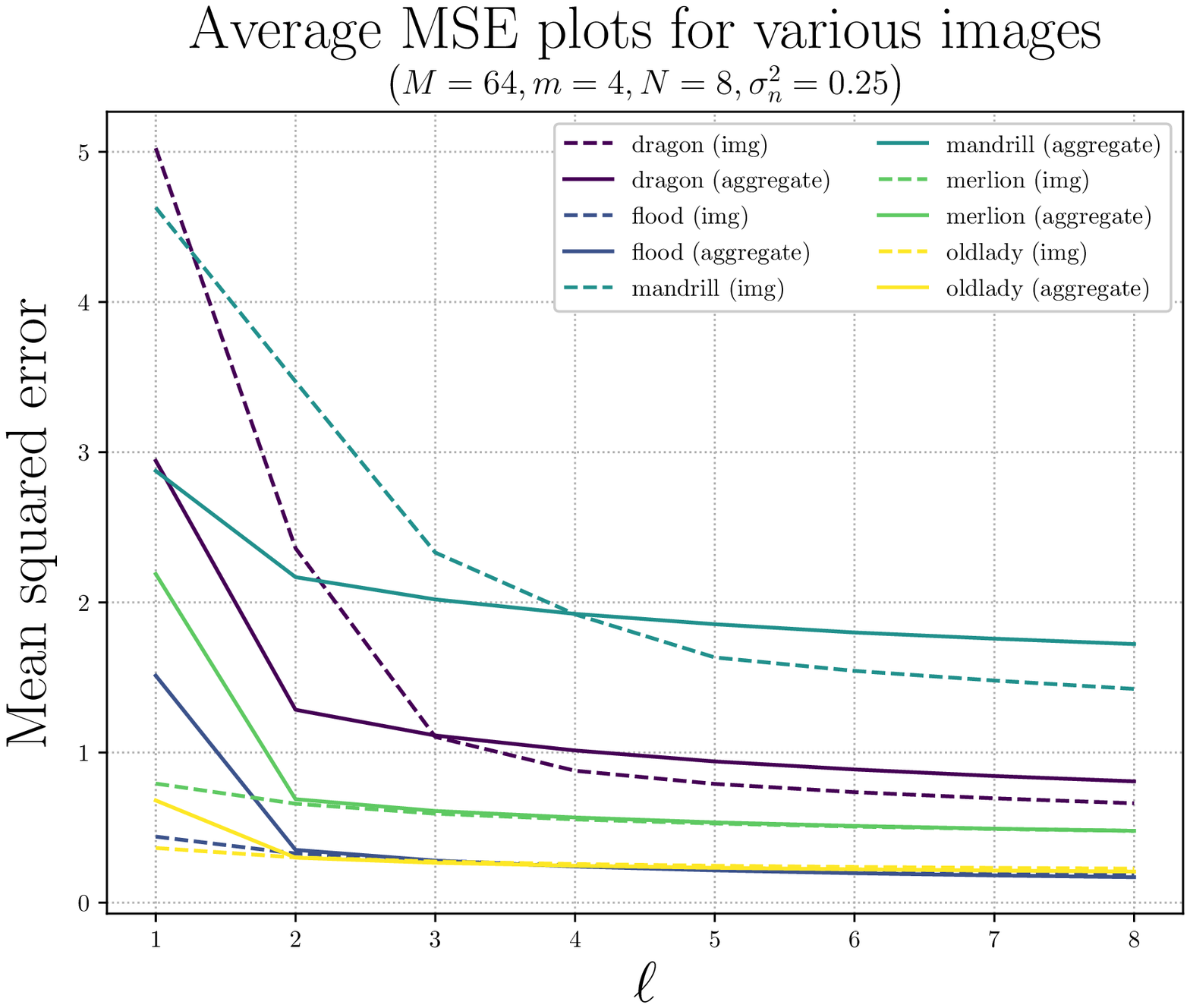} & 
	\includegraphics[width=0.47\linewidth]{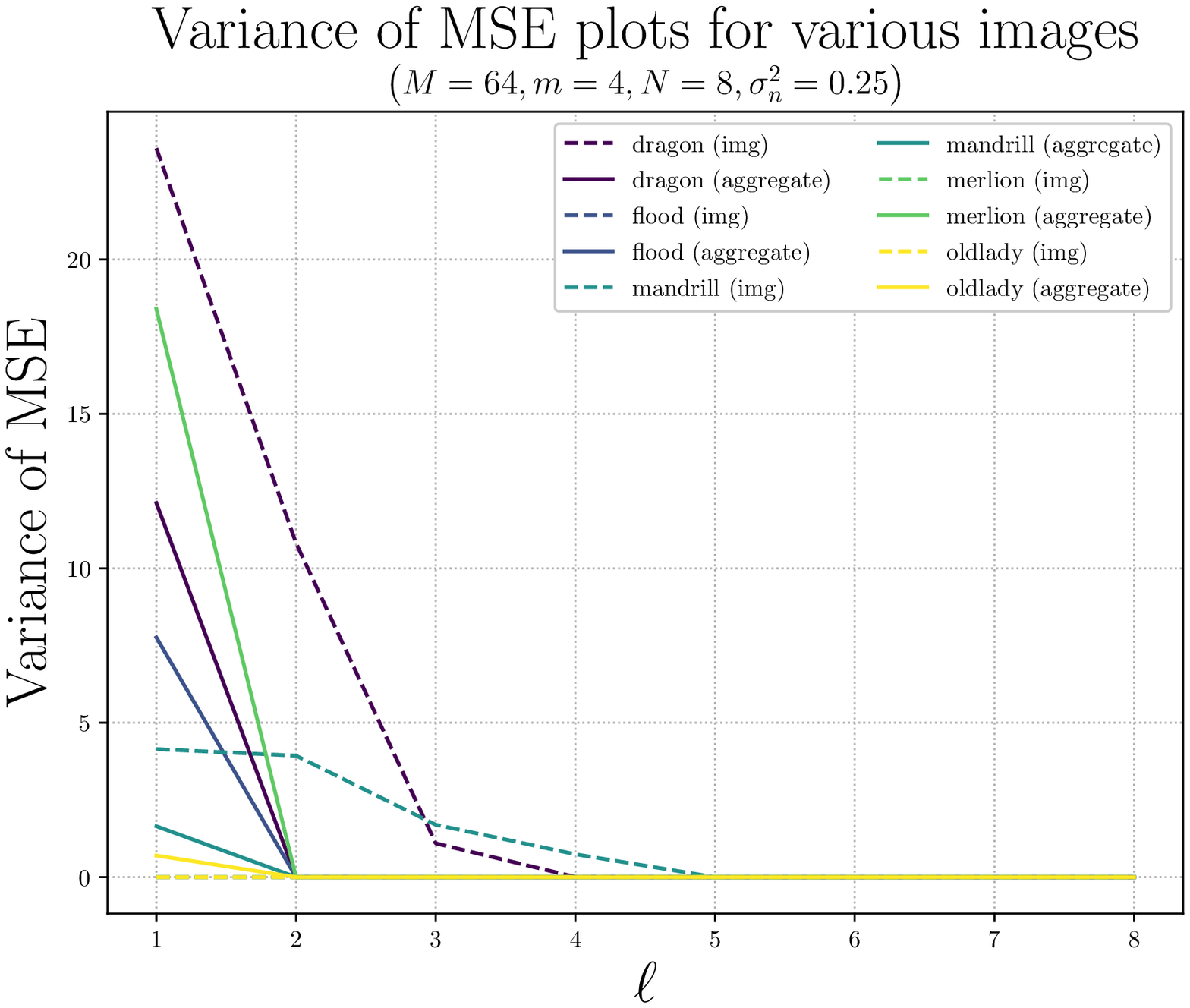} \\
	(a) Average of Incremental $\MSE$ & 
	(b) Variance in Incremental $\MSE$ \\
		
	\includegraphics[width=0.47\linewidth]{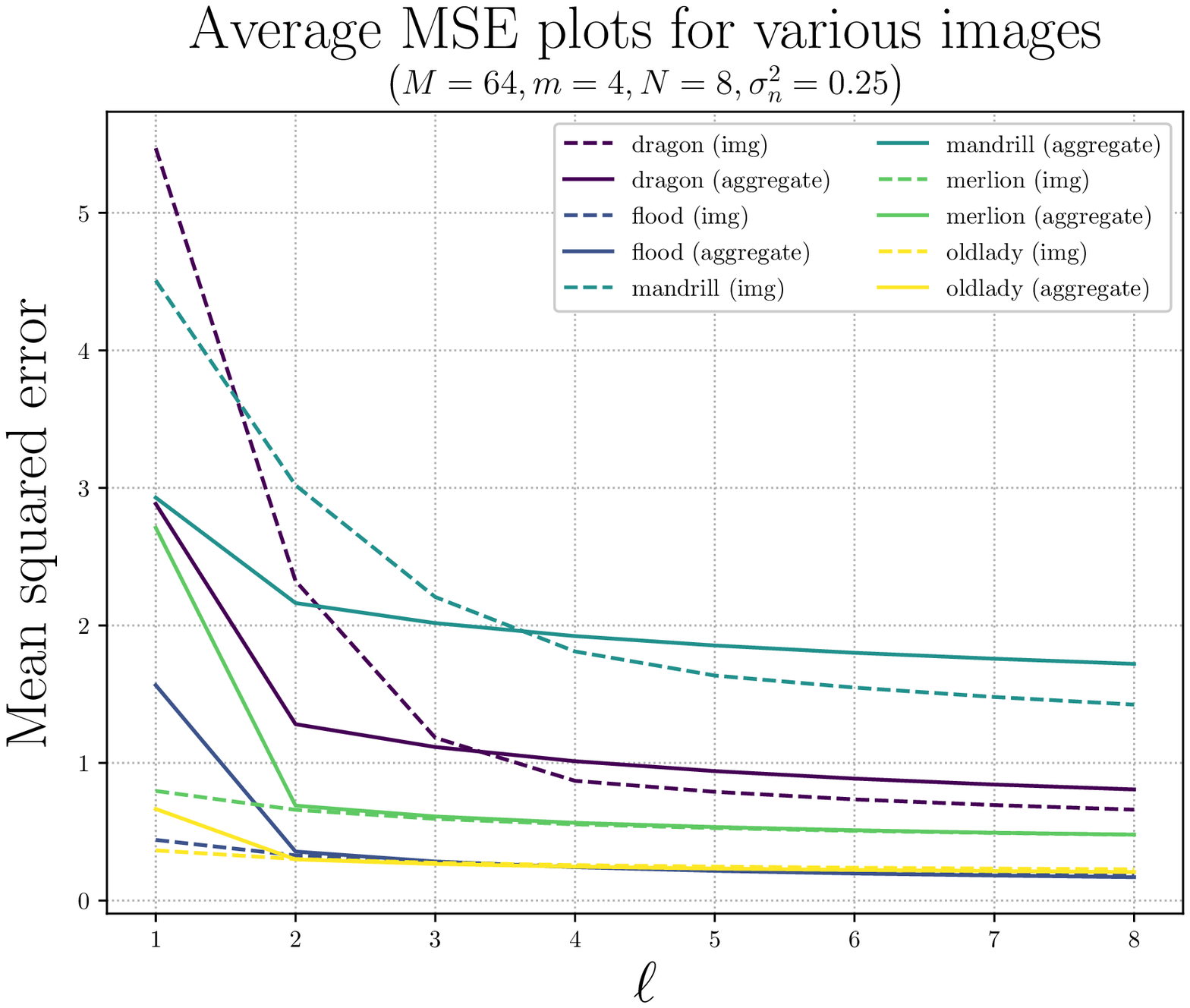} & 
	\includegraphics[width=0.47\linewidth]{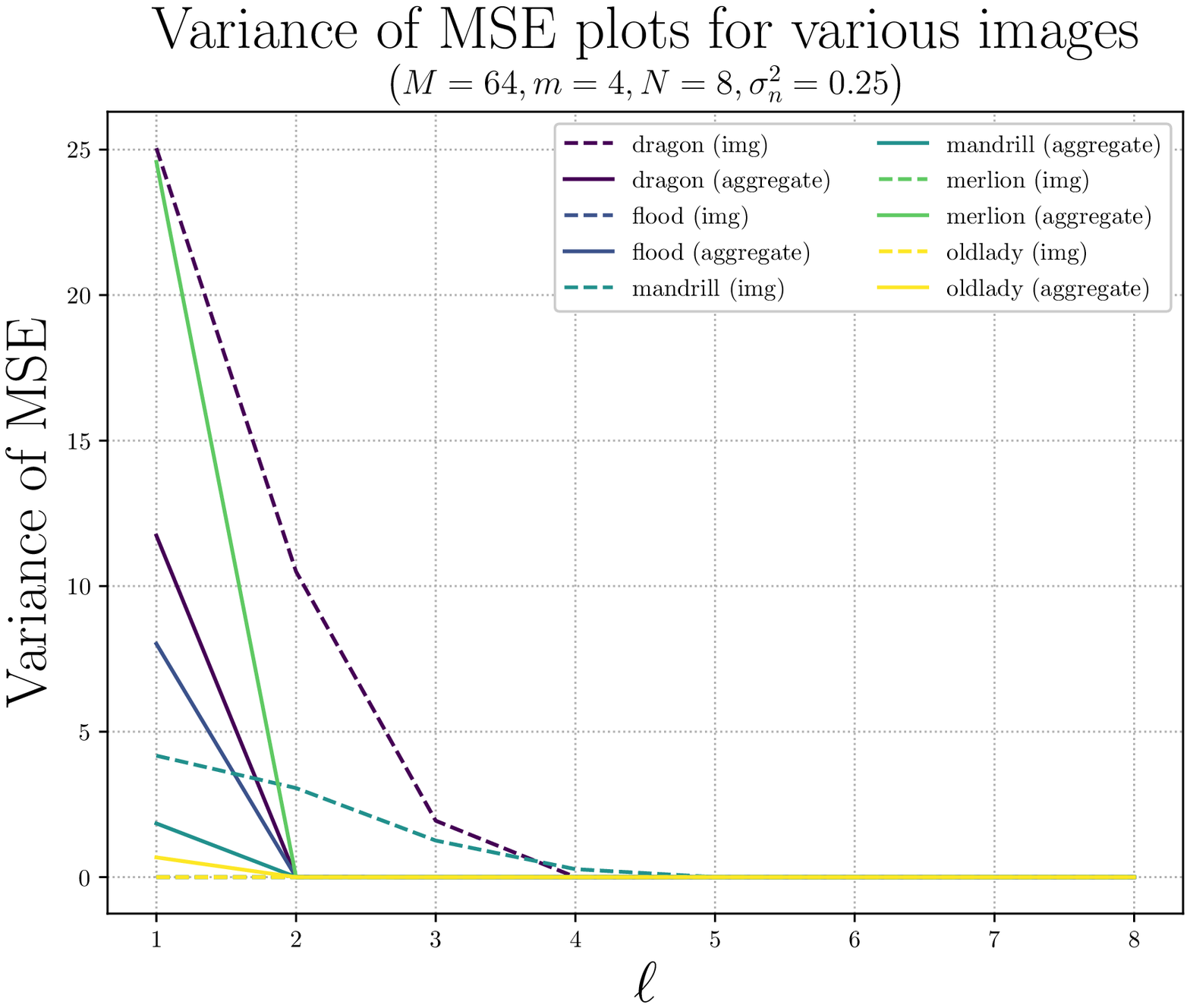} \\
	(c) Average of Randomized $\MSE$ & 
	(d) Variance in Randomized $\MSE$ \\
\end{tabular}
\caption{The average $\MSE$ plots and their respective variances obtained from recovery simulations based on two options of second order statistics, namely the {\tt aggregate} (aggregate) and the individual image (img). The plots in the upper row is based on the incremental recovery while those in the lower row come from the randomized one. We use Mode $1$ with $(M,m,N,\sigma_n^2)=(64,4,8,0.25)$.}
\label{fig:perf}
\end{figure}

There are insights to gain from the average recovery performance. While Mode $1$ is the closest to the original design philosophy of reaching the best $\MSE$ when all $N$ packets are available, its recovery performance fluctuates rather widely for $\ell < \frac{N}{2}$. Figure~\ref{fig:perf} presents the average $\MSE$ and its variance, simulated on the indicated images, for both the incremental and randomized procedures when $(M, m,N,\sigma_n^2) = (64,4,8,0.25)$.  

\section{Recovery and Performance Analysis on Control Images}\label{sec:control}

\begin{figure}[h!]
	\centering
	\begin{tabular}{cc}
		\includegraphics[width=0.44\linewidth]{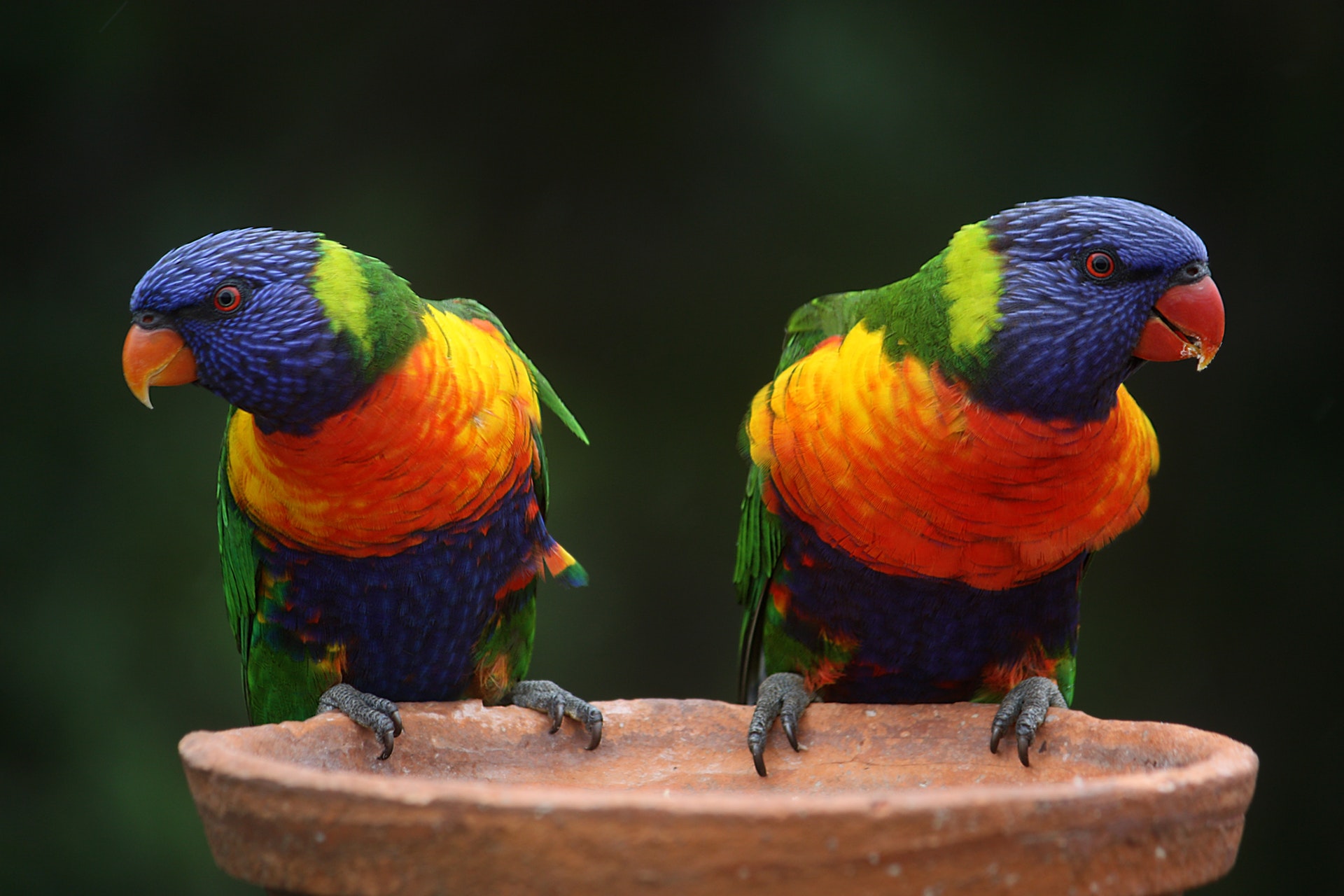} & 
		\includegraphics[width=0.44\linewidth]{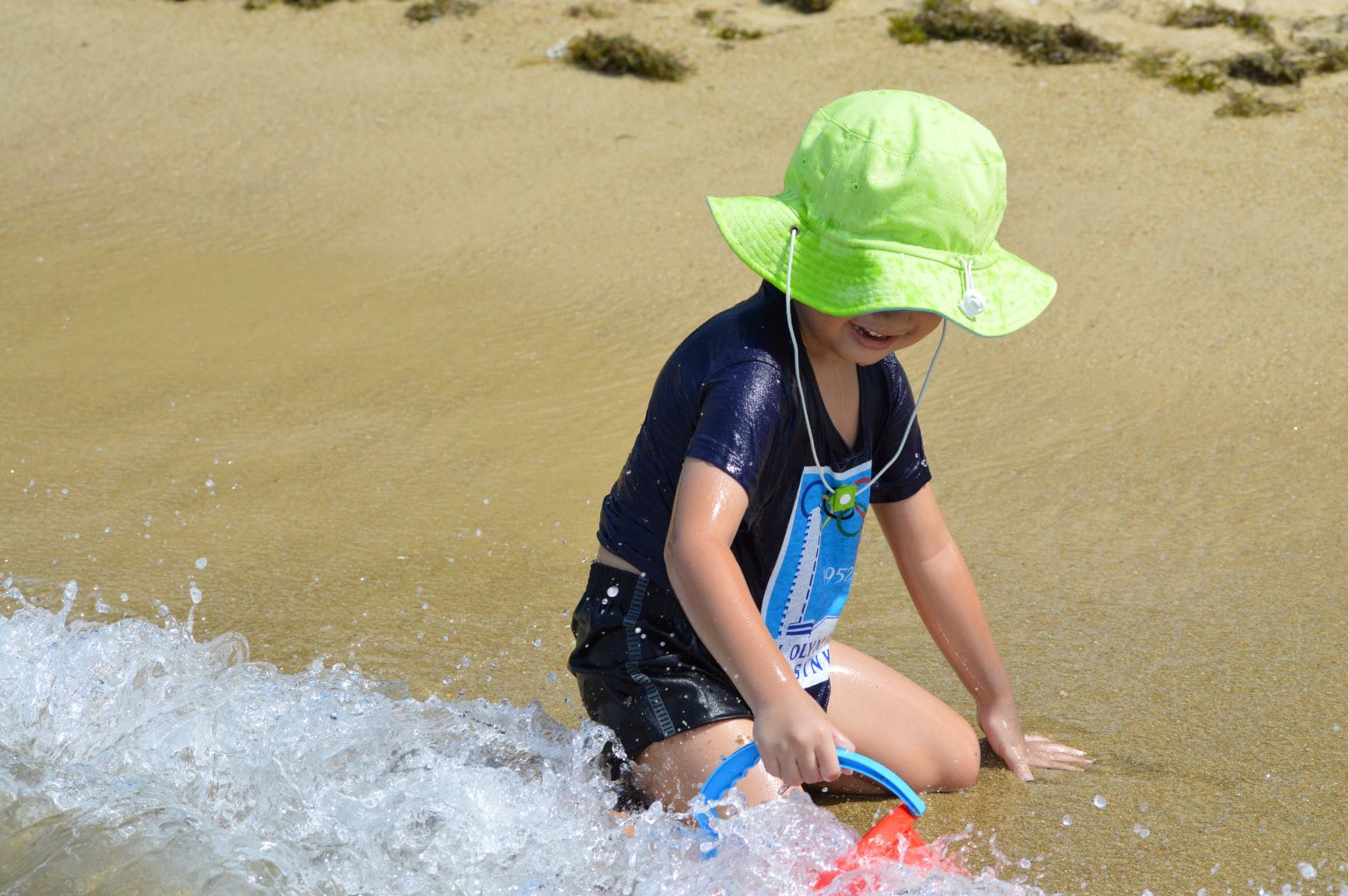} \\
		(a) {\tt twobirds} & (b) {\tt boy}
	\end{tabular}
	\caption{Two images from the control data set for illustrations in recovery and analysis. Taken from \url{https://pixabay.com/users/magee-830963/}, {\tt twobirds} depicts two rainbow lorikeets. The image {\tt boy} was taken by the corresponding author.}
	\label{fig:control}
\end{figure} 

\begin{figure}[h!]
\centering
\begin{tabular}{ccc}
\includegraphics[width=0.3\linewidth]{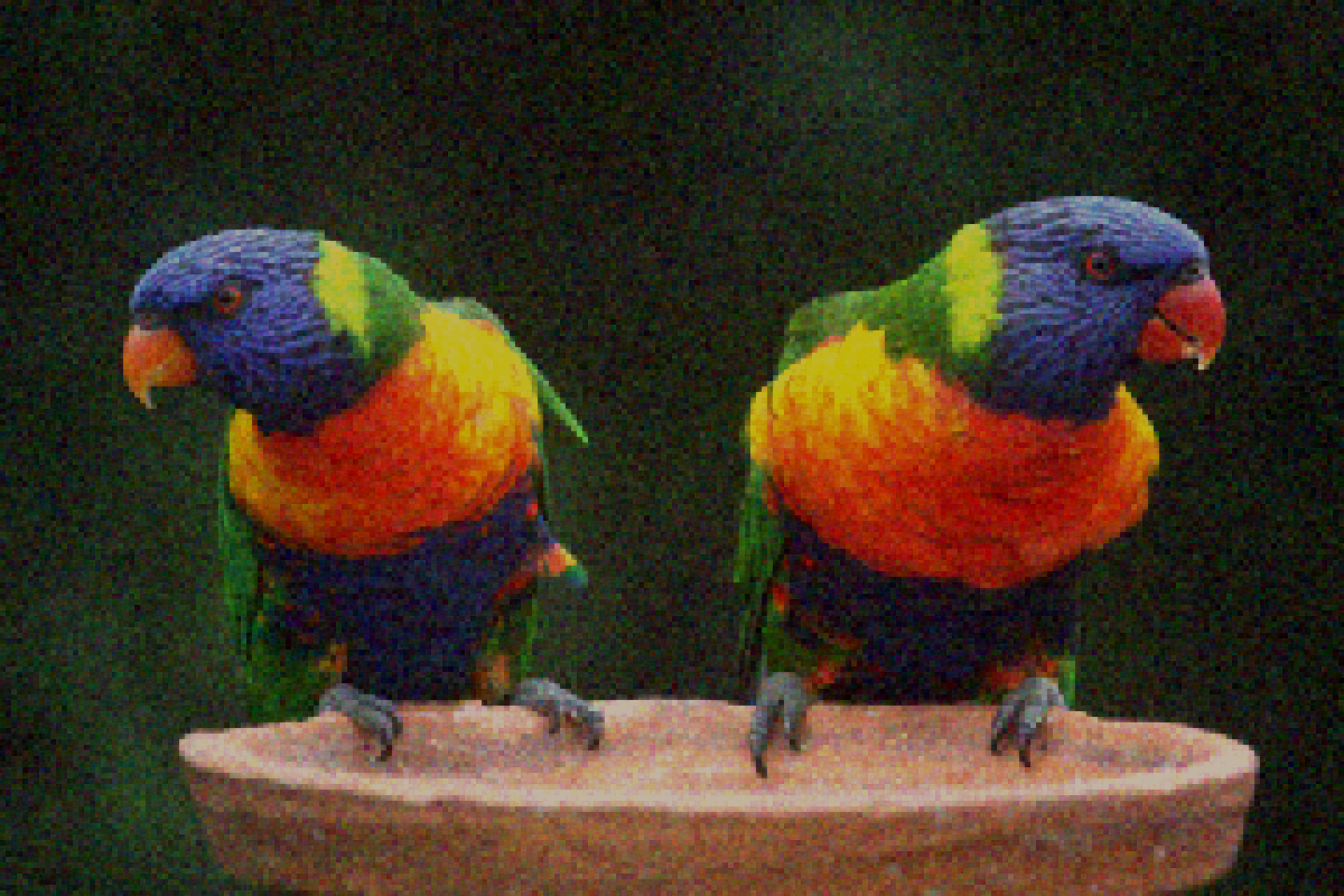} & 
\includegraphics[width=0.3\linewidth]{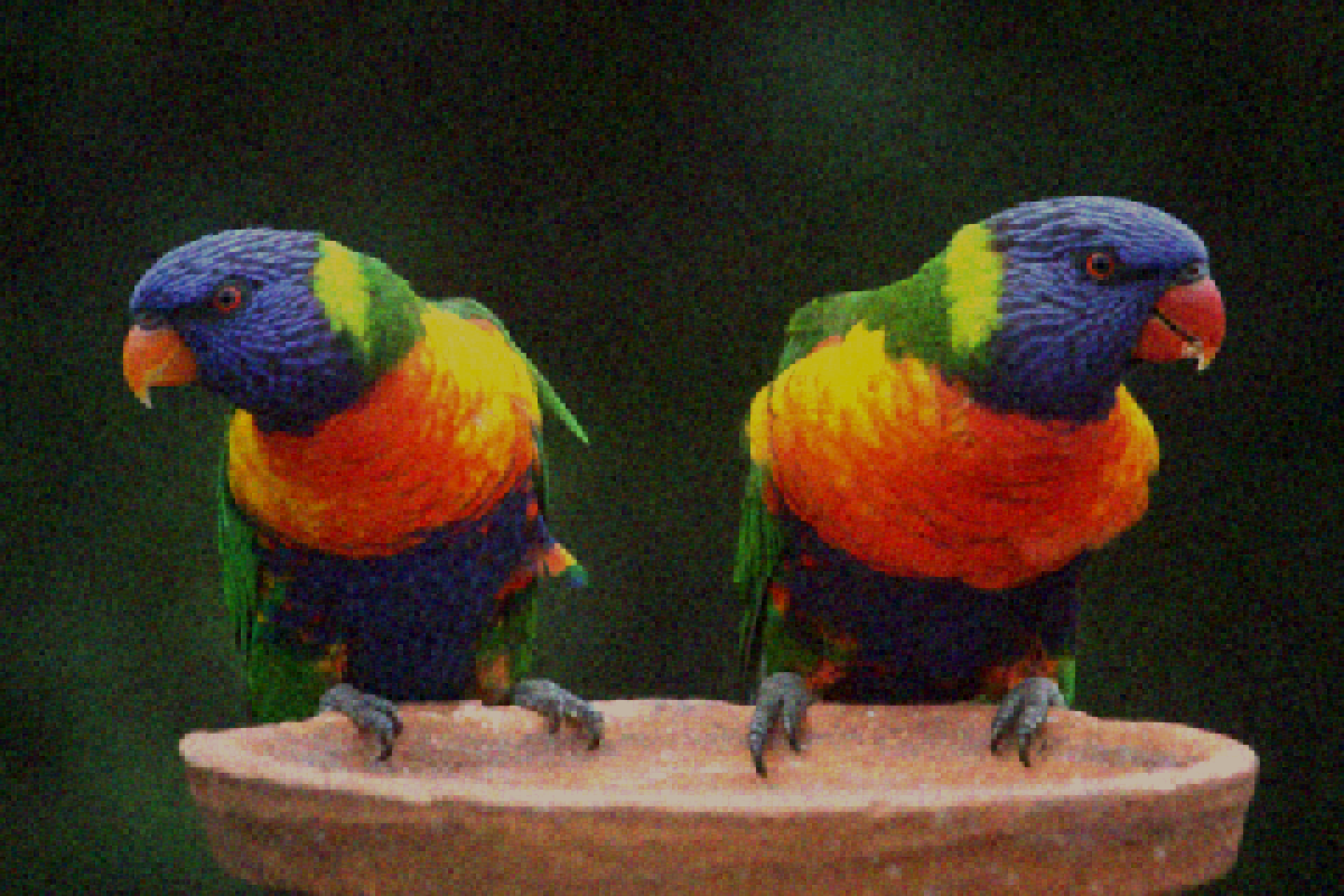} & 
\includegraphics[width=0.3\linewidth]{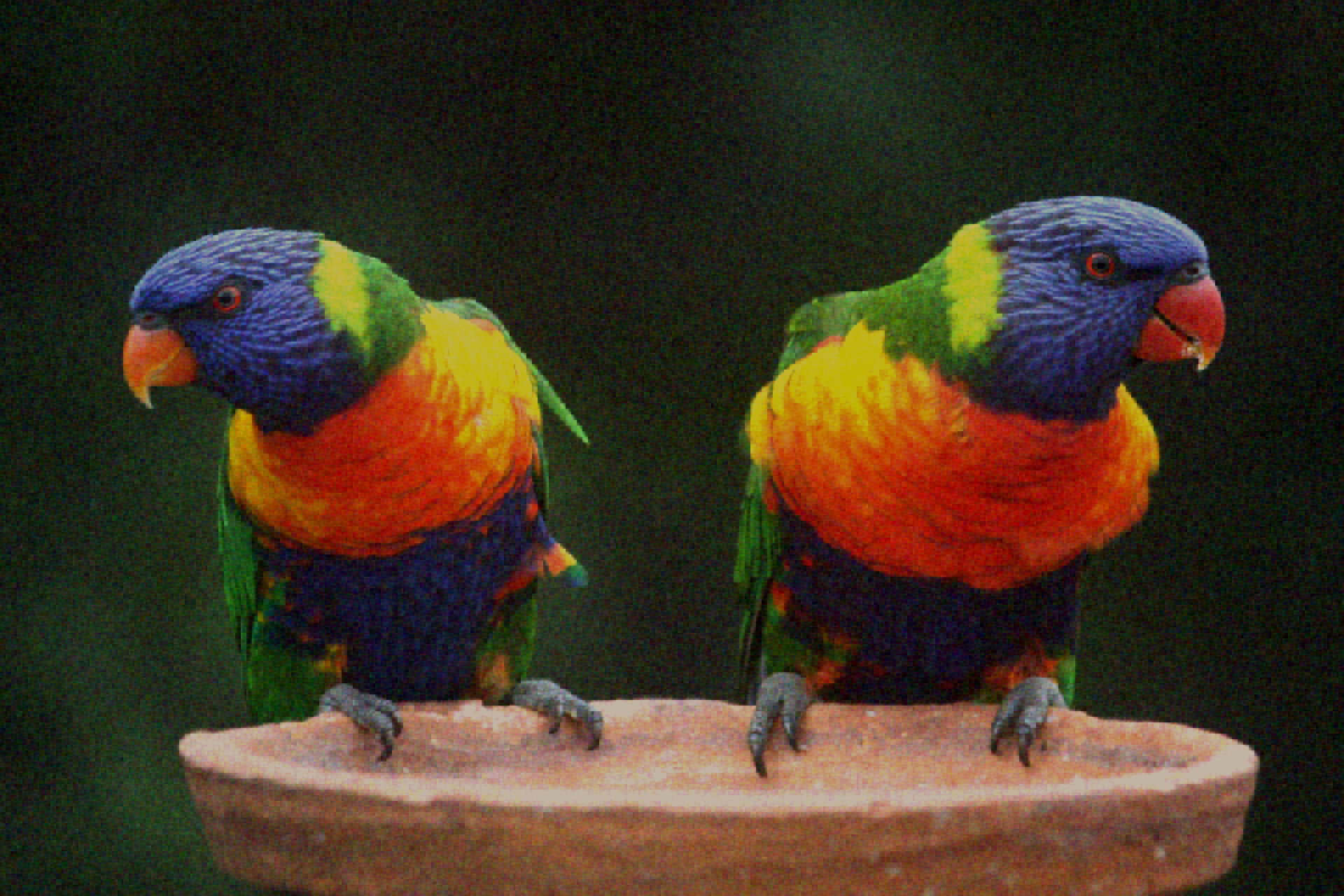} \\ 
(a) $\ell=1$ & (b) $\ell=2$ & (c) $\ell=4$\\
		
\includegraphics[width=0.3\linewidth]{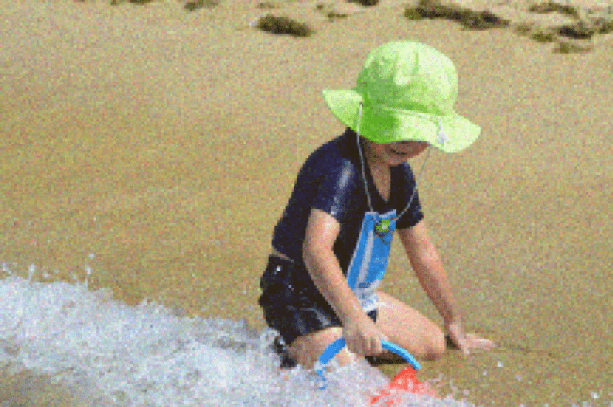} & 
\includegraphics[width=0.3\linewidth]{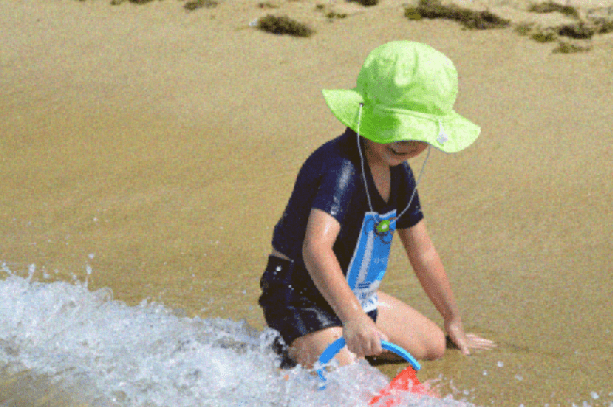} & 
\includegraphics[width=0.3\linewidth]{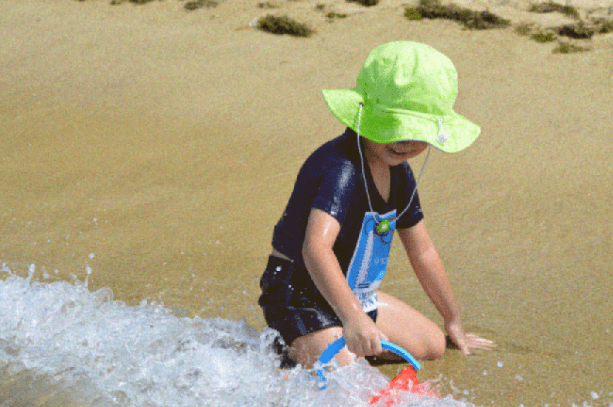} \\ 
(d) $\ell=1$ & (e) $\ell=4$ & (f) $\ell=8$ \\
\end{tabular}
\caption{Incrementally recovered {\tt twobirds} and {\tt boy}. On {\tt twobirds}, $(M,m,N,\sigma_n^2)=(64,8,8,0.64)$. By Table~\ref{table:distro} Entry 12, all modes have probing distribution $[8]^8$. Mode $2$ with $(M,m,N,\sigma_n^2)=(64,4,16,0.64)$ is used to recover {\tt boy}, where the probing distribution is as given in Table~\ref{table:distro} Entry 7.}
\label{fig:controlrec}
\end{figure}

In~\url{https://github.com/adamasstokhorst/holographic/tree/master/control} we collect some natural images for control purposes. They are sensed and recovered by using the $\Lambda$ of {\tt aggregate} obtained in Section~\ref{sec:comp}, since their second order statistics are assumed to be unknown. We use two control images, namely {\tt twobirds} and {\tt boy}, in Figure~\ref{fig:control} to illustrate and analyze their recovery. Their respective incrementally recovered images, for the indicated parameters, can be found in Figure~\ref{fig:controlrec}.

\begin{figure}[h!]
\centering
\begin{tabular}{cc}
	\includegraphics[width=0.44\linewidth]{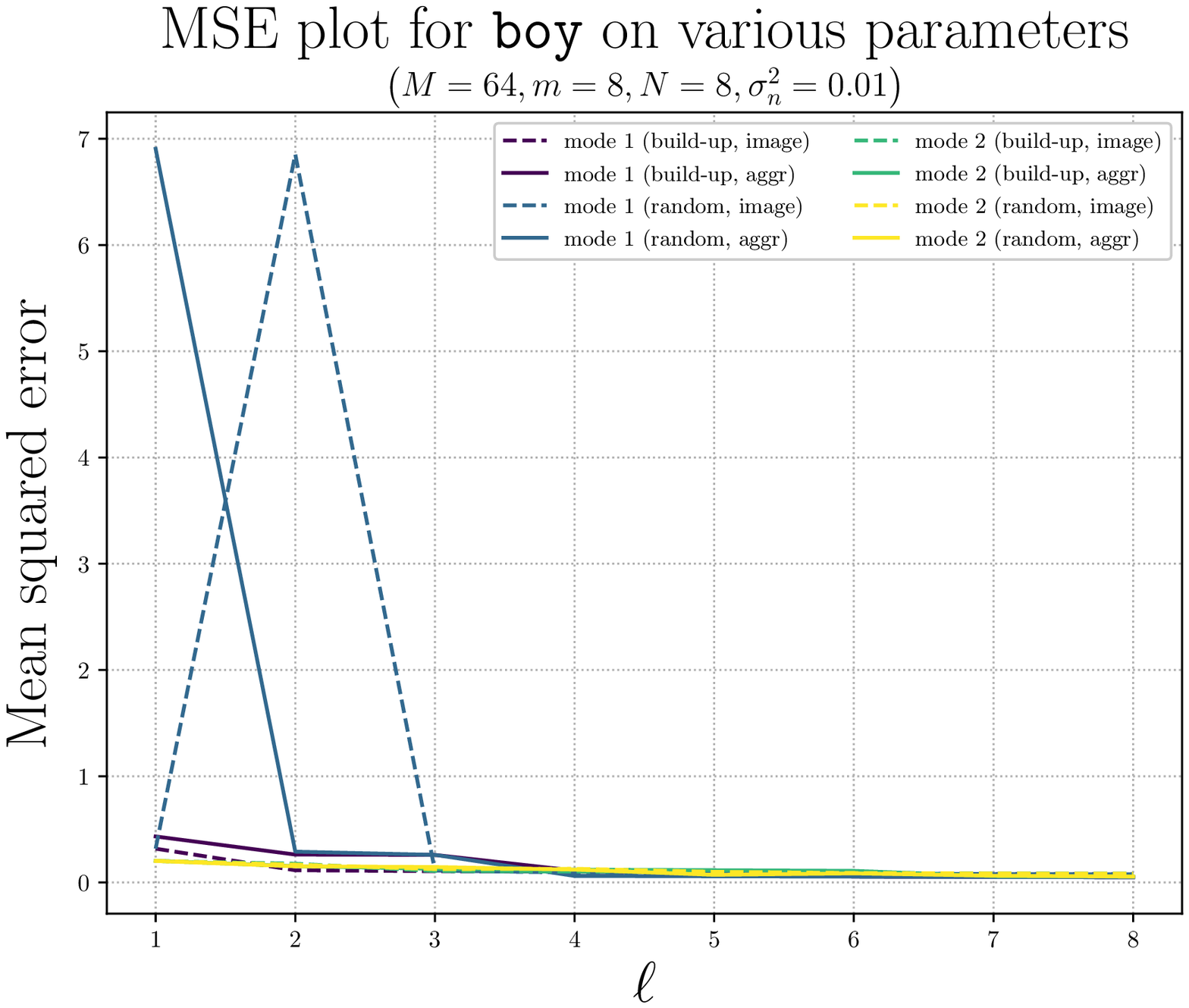} & 
	\includegraphics[width=0.44\linewidth]{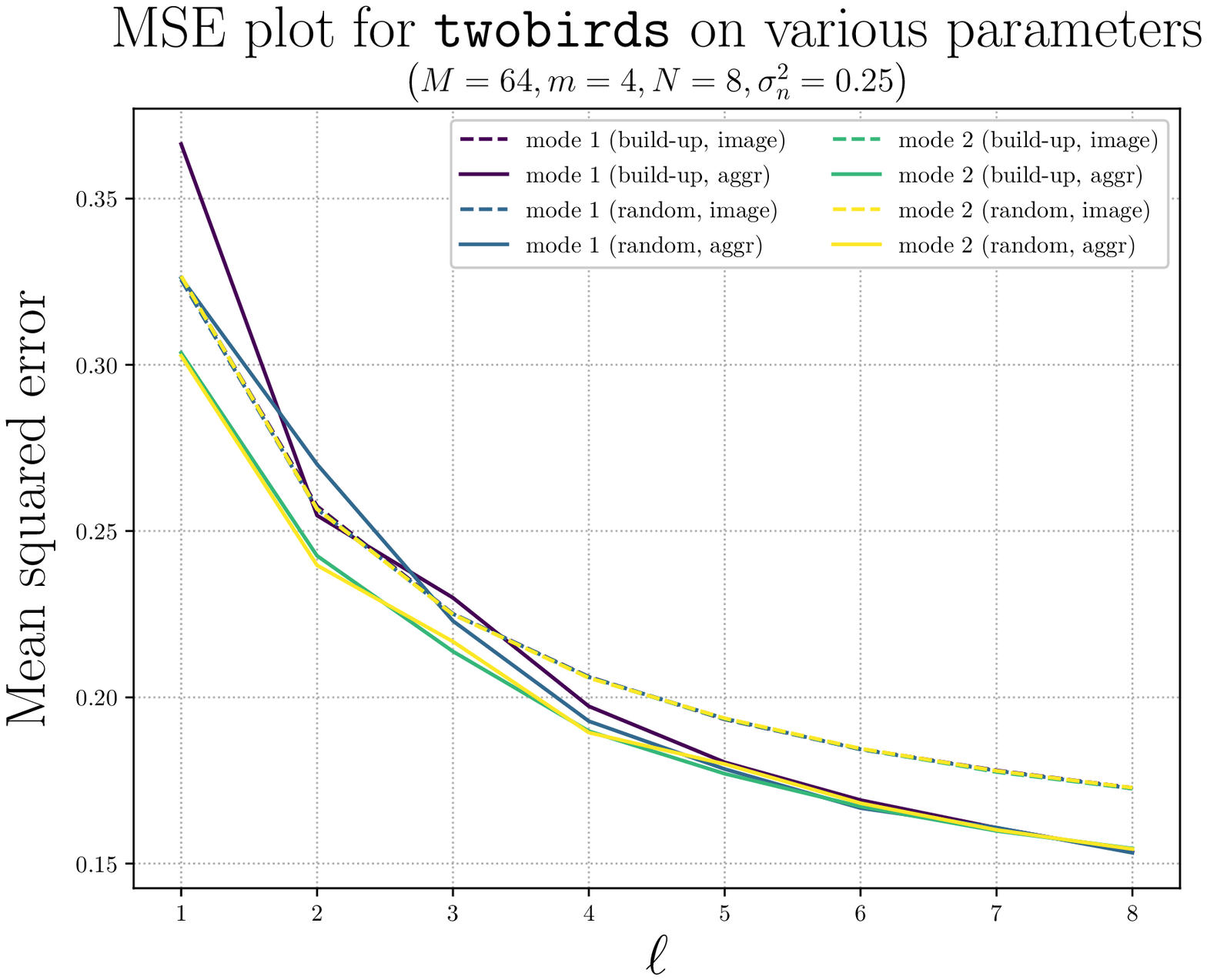} \\
	(a) Recovery instances on {\tt boy} & (b) Recovery instances on {\tt twobirds}
\end{tabular}
\caption{The $\MSE$ plots capture the performance of a single recovery run each on Modes $1$ and $2$, for the indicated settings, on {\tt boy} and {\tt twobirds}. We use both the $\Lambda$ profile of {\tt aggregate} (aggr) and that of the image, as well as both the incremental (build-up) and randomized (random) procedure.}
\label{fig:ctrl}
\end{figure}

We see in Figure~\ref{fig:ctrl} how the recovered images exhibit the expected properties of progressive refinement as more and more packets are used. The plots confirm that using the $\Lambda$ of {\tt aggregate} on control images results in a very similar $\MSE$ values as using the individual image's profile, had it been known and used.

\section{Concluding Remarks}\label{sec:conclusion}

We briefly reviewed the theory of holographic sensing and then explained in detail the process of designing the sensing packets before presenting the expected performance of the distributed holographic encodings of databases of images. We have outlined the full design of the system implementing the idea of holographic representations and tested the system's performance on a data set of $49$ natural images. This data set can be easily enlarged into a big database if desired. Similarly with the types of images. The system has been demonstrated to also perform well on randomly selected control images not present in the data set.

\begin{figure}[h!]
	\centering
	\includegraphics[width=0.60\linewidth]{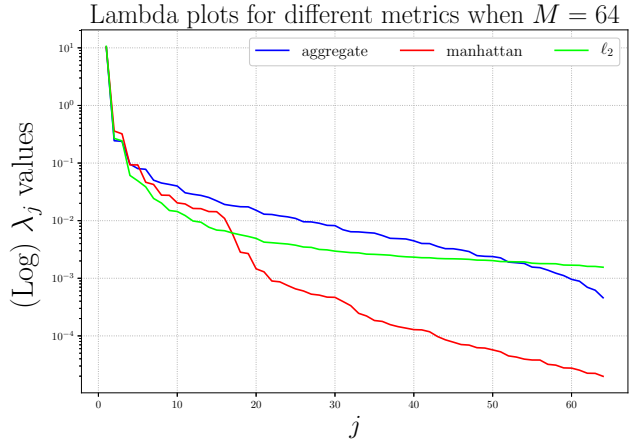}
	\caption{Comparison of the $\Lambda$ profiles of {\tt aggregate} and of the two models for $M=64$.}
	\label{fig:lambda_comp}
\end{figure}

\begin{figure}[h!]
\centering
\begin{tabular}{cc}
	\includegraphics[width=0.4\linewidth]{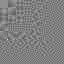} & 
	\includegraphics[width=0.4\linewidth]{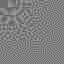} \\
	(a) $\Psi$ in {\tt grid} model & (b) $\Psi$ in {\tt line} model
\end{tabular}
\caption{The visualized $\Psi$ matrices to be compared with the one for {\tt aggregate} in Figure~\ref{fig:visual}.}
\label{fig:comp_model}
\end{figure}

In conclusion we also wish to mention two distance-based correlation models that work quite well in estimating $\BR_{xx}$ for natural images. We call them the {\tt grid} model, when the distance is the Manhattan length, and the {\tt line} model, when the $\ell_2$-norm is used. The respective models can be formulated as follows. Let $\BR^{\mathrm{G}}$ and $\BR^{\mathrm{L}}$ be the respective $M \times M$ matrices that approximate $\BR_{xx}$ in the {\tt grid} and {\tt line} models. Without loss of generality, we can assume $M$ to be a square, \eg, $8^2$ as in our typical implementation, and let $\nu :=\sqrt{M}$. Since correlation decays as a function of distance, the entries in the model matrices must faithfully reflect this fact. Let $i := \nu \cdot a + b$ and $j:= \nu \cdot p + q$. For $i,j \in \bbra{M}$, let 
\begin{equation}\label{eq:matmod}
\BR^{\mathrm{G}}_{i,j} = A \cdot \gamma^{|a-p| + |b-q|} \mbox{ and }  
\BR^{\mathrm{L}}_{i,j} = A \cdot \gamma^{\sqrt{(a-p)^2 + (b-q)^2}}.
\end{equation}
One can then proceed to determine the values of the constants $A$ and $\gamma$ that make the respective matrices closely resemble some experimentally obtained entries of $\BR_{xx}$ for a class of images. For $M=64$, based on our input data set of images, we obtain $A=0.18$ and $\gamma = 0.98$. The $\Lambda$ profiles as well as the corresponding $\Psi$ matrices can then be computed as above. Figures~\ref{fig:lambda_comp} and~\ref{fig:comp_model} provide, respectively, a useful $\Lambda$ comparison plot and a visualization of each of the $\Psi$ matrices for inspection. They show that the {\tt line} model fits the values from our image data set better than the {\tt grid} model.

\begin{figure}[h!]
	\centering
	\begin{tabular}{cc}
		\includegraphics[width=0.44\linewidth]{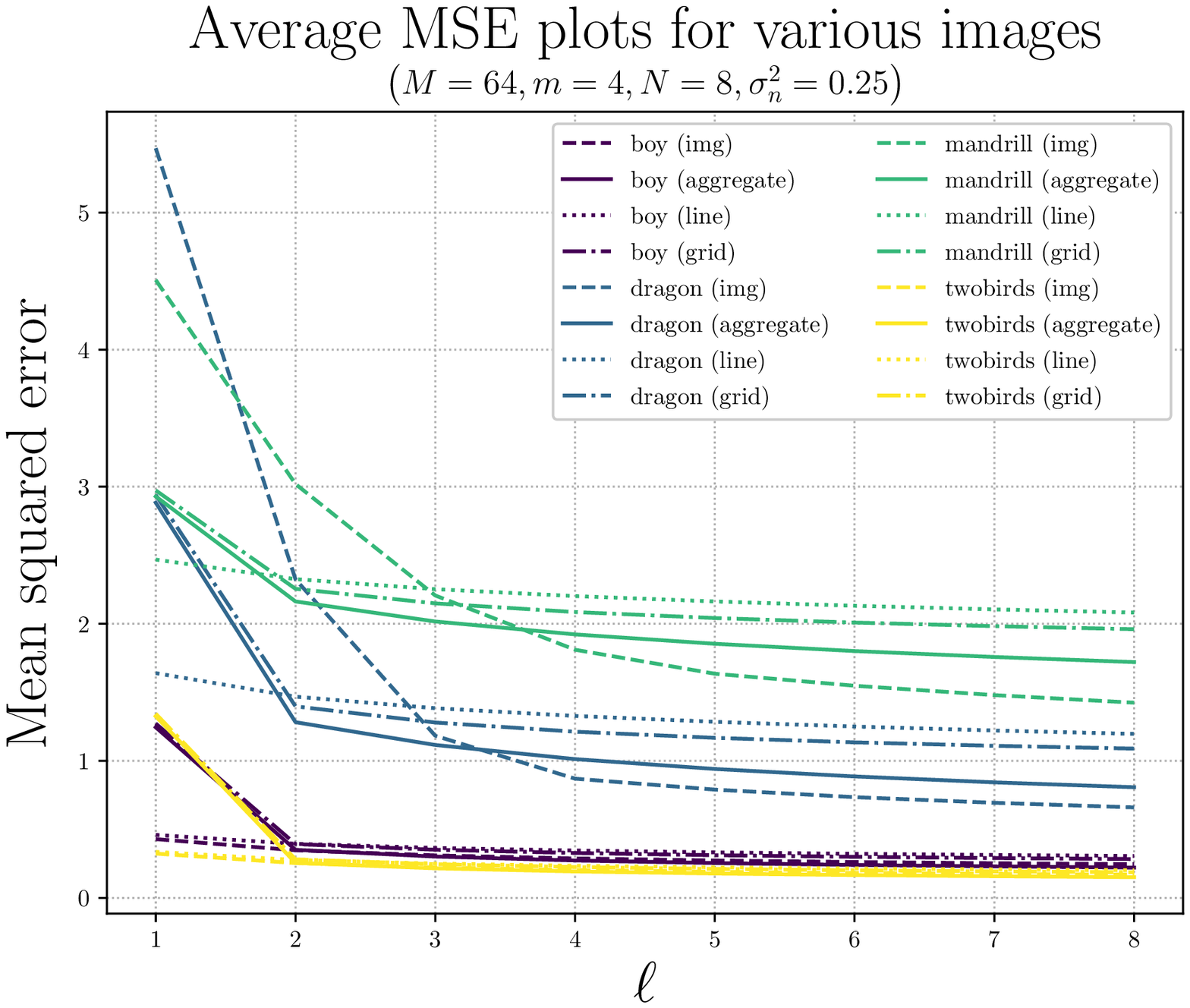} & 
		\includegraphics[width=0.44\linewidth]{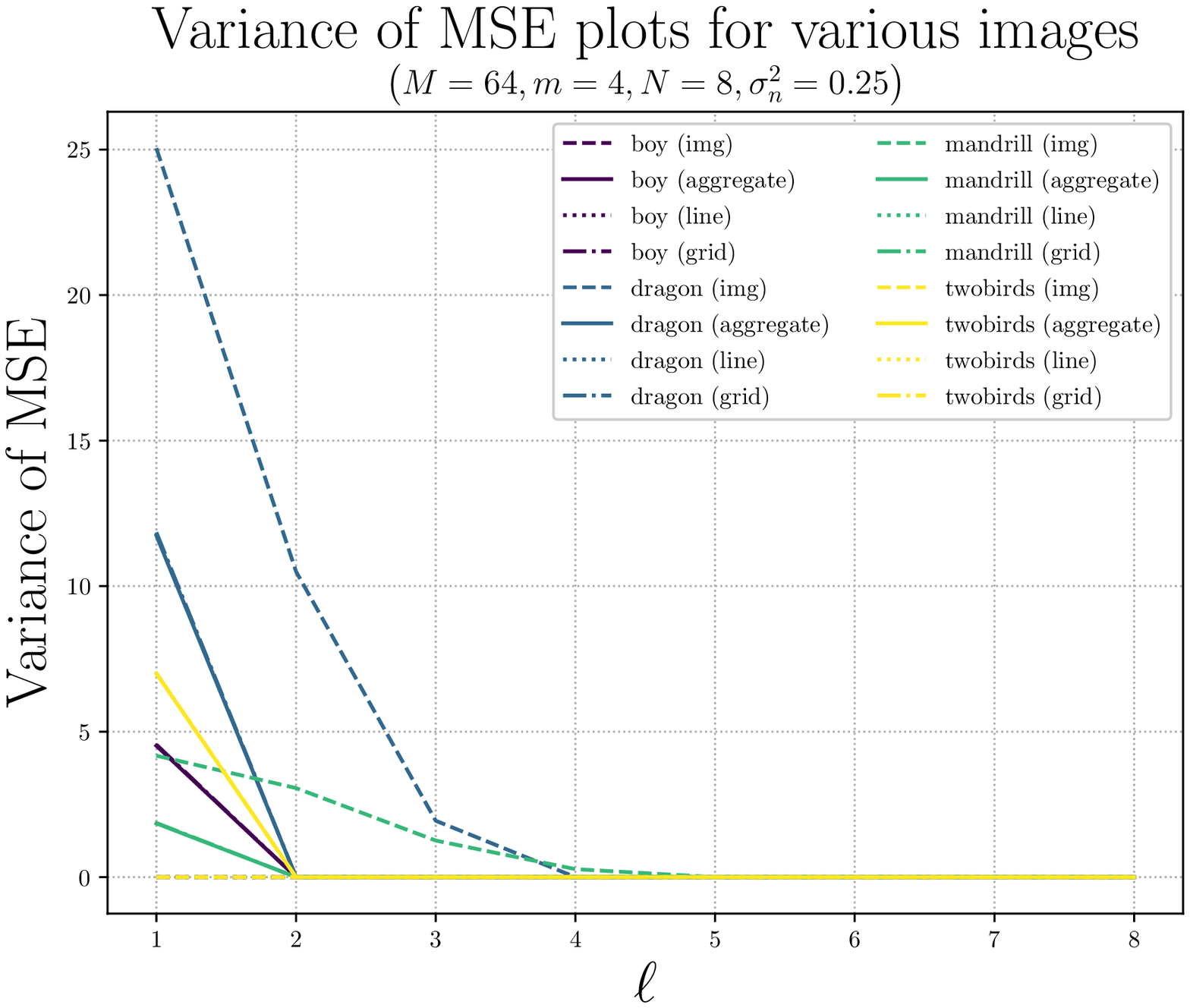} \\
		(a) Average of Randomized $\MSE$ & (b) Variance in Randomized $\MSE$\\
		\includegraphics[width=0.44\linewidth]{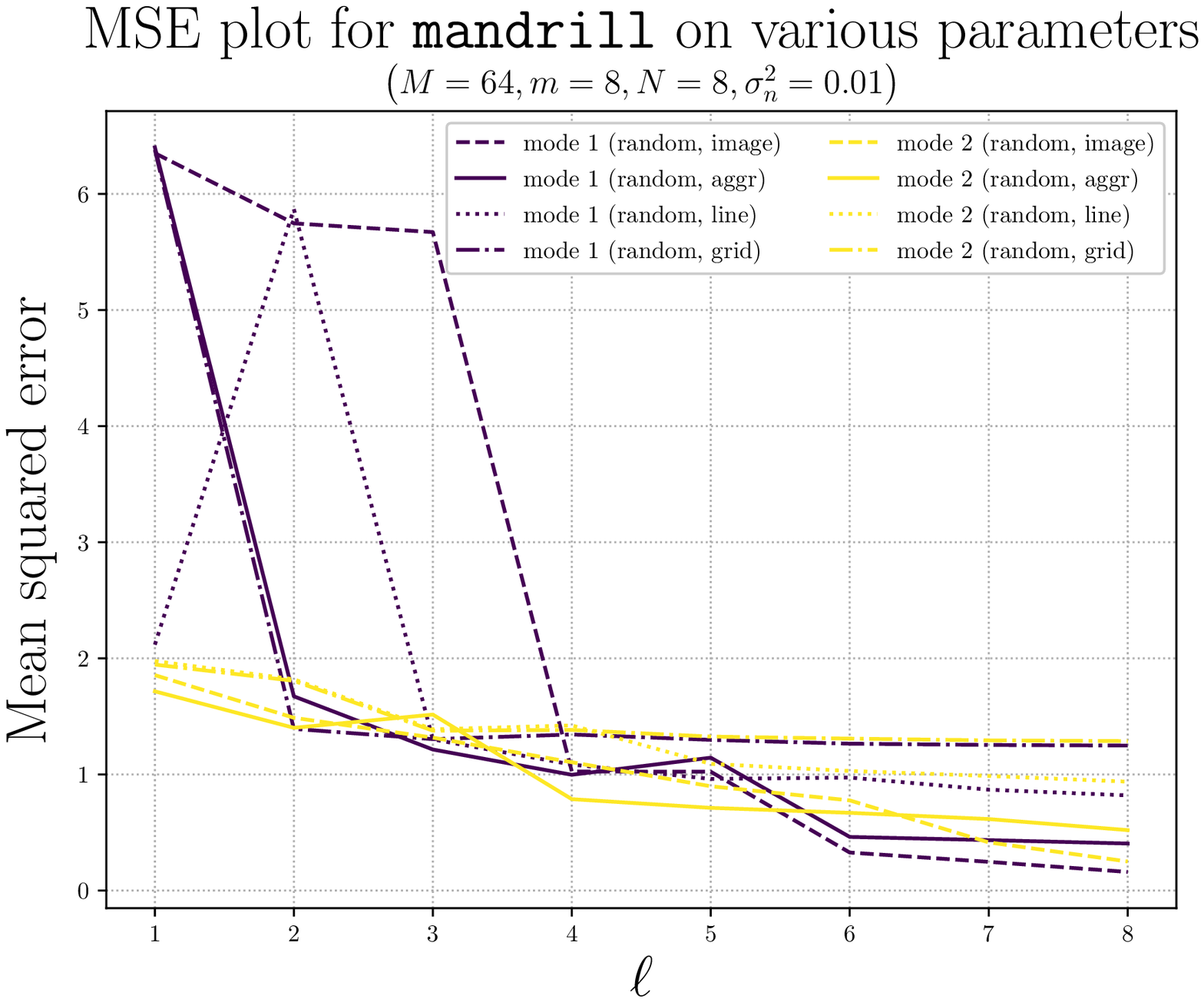} & 
		\includegraphics[width=0.44\linewidth]{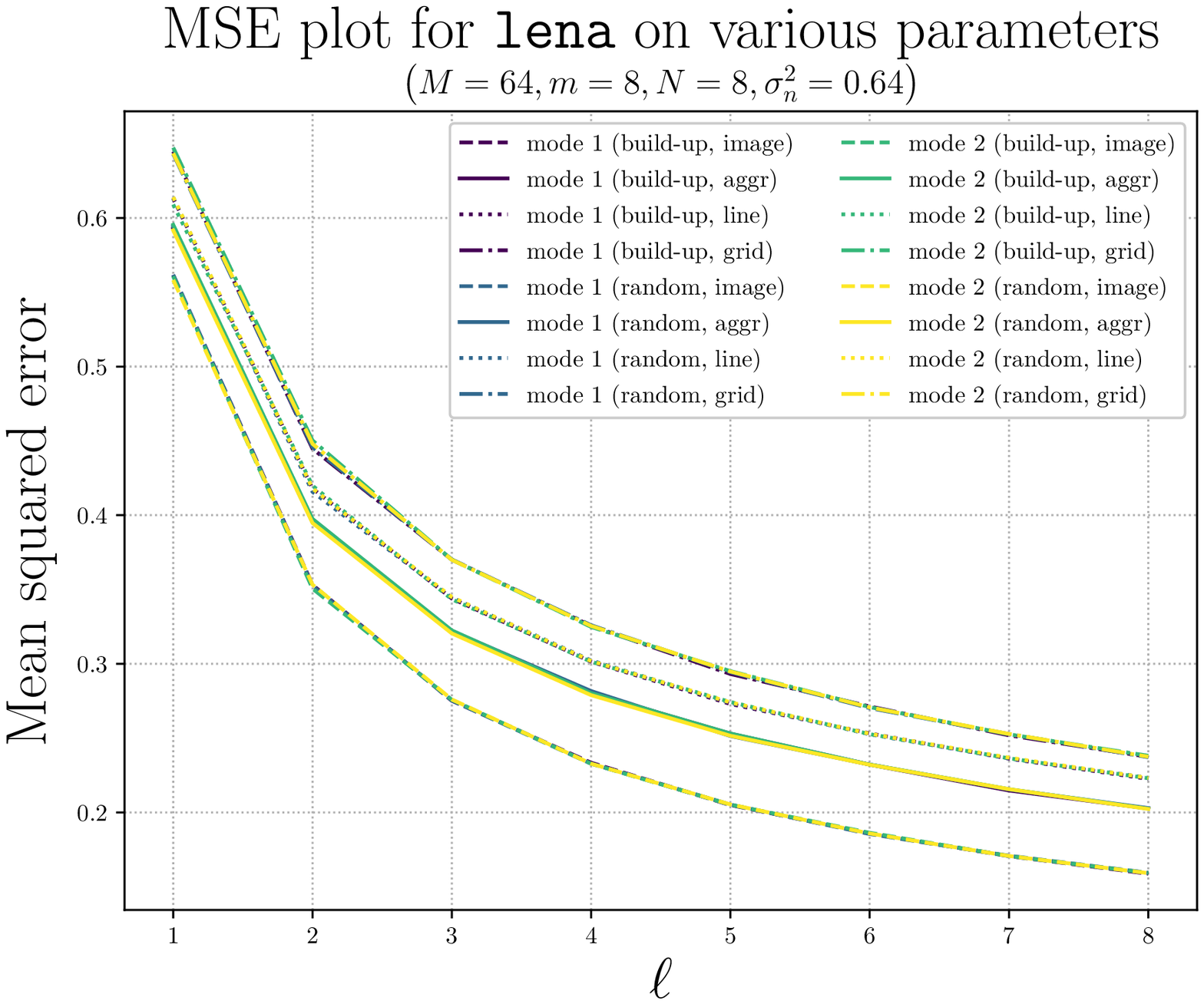} \\
		(c) Performance on {\tt mandrill} & (d) Performance on {\tt lena}
	\end{tabular}
	\caption{The first row displays the \emph{average} $\MSE$ plots for the \emph{randomized} recovery simulations on {\tt boy}, {\tt dragon}, {\tt mandrill}, and {\tt twobirds} on Mode $1$ with $(M,m,N,\sigma_n^2)=(64,4,8,0.25)$. The second row displays the performance on a single recovery instance for the indicated settings on {\tt mandrill} and {\tt lena}. Notice that the captured randomized recovery instance on Mode $1$ based on the $\Lambda$ of {\tt mandrill} is particularly bad. While such an instance is unlikely to occur, it highlights the potential drawback of using the randomized procedure in Mode $1$ on some images. Taken together, the plots allow for a quick comparison on the recovery performance across the statistical properties of a specific image, of {\tt aggregate}, and in the {\tt grid} and {\tt line} models, for various settings.}
	\label{fig:comp_rec}
\end{figure}

Using a reasonably fitting model may remove the need to compute for the second order statistics from some database prior to deployment. Figure~\ref{fig:comp_rec} shows that there is indeed not much of a difference in the recovery performance, either on a single recovery run or on average. 

\section*{Acknowledgments}
We thank Professor Vivek Goyal and the referees for their valuable comments and suggestions that helped us improve the overall presentation of this paper.


\end{document}